\documentclass[aps,pre,showpacs,amsmath,amssymb,amsfonts,superscriptaddress,twocolumn,longbibliography,preprintnumbers]{revtex4-1}

\usepackage{aas_macros}
\usepackage{latexsym}
\usepackage[caption=false]{subfig}
\usepackage{feyn}
\usepackage{feynmp}
\usepackage[toc,page]{appendix}
\usepackage{graphicx}
\usepackage[colorlinks=true,citecolor=blue,linkcolor=blue,urlcolor=blue]{hyperref}

\newcommand{\tr}[1]{\mathrm{tr}\left\{#1\right\}}
\newcommand{\la}{\left\langle}
\newcommand{\ra}{\right\rangle}
\newcommand{\e}[1]{\exp{\left(#1\right)}}
\newcommand{\be}{\begin{equation}}
\newcommand{\ee}{\end{equation}}
\newcommand{\eins}{\mbox{$1 \hspace{-1.0mm} {\bf l}$}}
\newcommand{\doubleoverarrow}[1]{\overset{\text{\tiny$\leftrightarrow$}}{#1}}
\newcommand{\mc}[1]{\mathcal{#1}}
\newcommand{\mrm}[1]{\mathrm{#1}}

\DeclareMathOperator*{\SumInt}{
\mathchoice
  {\ooalign{$\displaystyle\sum$\cr\hidewidth$\displaystyle\int$\hidewidth\cr}}
  {\ooalign{\raisebox{.14\height}{\scalebox{.7}{$\textstyle\sum$}}\cr\hidewidth$\textstyle\int$\hidewidth\cr}}
  {\ooalign{\raisebox{.2\height}{\scalebox{.6}{$\scriptstyle\sum$}}\cr$\scriptstyle\int$\cr}}
  {\ooalign{\raisebox{.2\height}{\scalebox{.6}{$\scriptstyle\sum$}}\cr$\scriptstyle\int$\cr}}
}

\begin{document}

\preprint{CALT-TH-2017-052}

\title{Jarzynski Equality for Driven Quantum Field Theories}

\author{Anthony Bartolotta}
\email{abartolo@theory.caltech.edu}
\affiliation{Walter Burke Institute for Theoretical Physics, California Institute of Technology, Pasadena, CA 91125 USA}

\author{Sebastian Deffner}
\email{deffner@umbc.edu}
\affiliation{Department of Physics, University of Maryland Baltimore County, Baltimore, MD 21250 USA}

\date{\today}


\begin{abstract}
The fluctuation theorems, and in particular, the Jarzynski equality, are the most important pillars of modern non-equilibrium statistical mechanics. We extend the quantum Jarzynski equality together with the Two-Time Measurement Formalism to their ultimate range of validity -- to quantum field theories. To this end, we focus on a time-dependent version of scalar phi-four. We find closed form expressions for the resulting work distribution function, and we find that they are proper physical observables of the quantum field theory. Also, we show explicitly that the Jarzynski equality and Crooks fluctuation theorems hold at one-loop order independent of the renormalization scale. As a numerical case study, we compute the work distributions for an infinitely smooth protocol in the ultra-relativistic regime. In this case, it is found that work done through processes with pair creation is the dominant contribution.
\end{abstract}

\maketitle

\section{Introduction}

In physics there are two kinds of theories to describe motion: microscopic theories whose range of validity is determined by a length scale and the amount of kinetic energy, such as classical mechanics or quantum mechanics; and phenomenological theories, such as thermodynamics, which are valid as long as external observables remain close to some equilibrium value.

Over the last two centuries, microscopic theories have undergone a rapid development from classical mechanics over special relativity and quantum mechanics to quantum field theory. While quantum field theories were originally developed for particle physics and cosmology, this apporach has also been shown to be powerful in the description of condensed matter systems. Examples include quasiparticle excitations in graphene, cavity quantum electrodynamics, topological insulators, and many more \cite{Walther2006,Tsvelik2007,Nagaosa2013,Fradkin2013}. 

In contrast to the evolution of microscopic theories, the development of thermodynamics has been rather stagnant -- until only two decades ago when the first fluctuation theorems were discovered \cite{PhysRevLett.71.2401, PhysRevLett.71.3616, PhysRevLett.74.2694,  PhysRevE.60.159}. Conventional thermodynamics can only fully describe infinitely slow, equilibrium processes. About all real, finite-time processes the second law of thermodynamics only asserts that some amount of entropy is dissipated into the environment, which can be expressed with the average, irreversible entropy production as $\la\Sigma\ra\geq0$ \cite{Callen1985}. The (detailed) fluctuation theorem makes this statement more precise by expressing that negative fluctuations of the entropy production are exponentially unlikely \cite{PhysRevLett.71.2401, PhysRevLett.71.3616, PhysRevLett.74.2694,  PhysRevE.60.159, Crooks1999}, 
\begin{equation}
\label{eq00}
\mc{P}(-\Sigma)=\e{-\Sigma}\,\mc{P}(\Sigma)\,. 
\end{equation}

The most prominent (integral) fluctuation theorem \cite{OrtizdeZarate2011} is the Jarzynski equality \cite{Jarzynski1997}, which holds for all systems initially prepared in equilibrium and undergoing isothermal processes,
\begin{equation}
\label{eq01}
\la\e{-\beta W}\ra=\e{-\beta \Delta F}\,,
\end{equation}
where $\beta$ is the inverse temperature, $W$ is the thermodynamic work, and $\Delta F$ is the free energy difference between the instantaneous equilibrium states at the initial and final times. In its original inception the Jarzynski equality \eqref{eq01} was formulated for classical systems with Hamiltonian \cite{Jarzynski1997} and Langevin dynamics \cite{Jarzynski1997PRE}. Thus, $W$ is essentially a notion from classical mechanics, where work is given by a force along a trajectory. The advent of modern fluctuation theorems for classical systems \cite{PhysRevLett.71.2401, PhysRevLett.71.3616, PhysRevLett.74.2694, Jarzynski1997, Jarzynski1997PRE,PhysRevE.60.159, PhysRevLett.95.040602,Crooks1999} has spurred the development of a new field, which has been dubbed stochastic thermodynamics \cite{Broeck1986,0034-4885-75-12-126001, doi:10.1146/annurev-conmatphys-062910-140506,  doi:10.1146/annurev.physchem.58.032806.104555, doi:10.1080/00018730210155133, PhysRevX.7.021051}.

In the study of nanoscale systems out of thermal equilibrium, it is natural to ask in what regimes quantum effects become significant and how  fluctuation theorems apply to quantum systems \cite{PhysRevE.71.066102, PhysRevLett.90.170604, 0295-5075-79-1-10003,Hanggi2015, RevModPhys.83.771, RevModPhys.83.1653,RevModPhys.81.1665, RevModPhys.86.1125, PhysRevE.73.046129, PhysRevLett.92.230602, PhysRevE.72.027102, 1751-8121-40-26-F08, PhysRevE.77.051131,Allahverdyan2014}. Nevertheless, it took another decade before it was clearly stated that in quantum mechanical systems $W$ is not a quantum observable in the usual sense \cite{PhysRevE.75.050102}. This means that there is no hermitian operator, whose eigenvalues are the classically observable values of $W$. This is the case because thermodynamic work is a path dependent quantity -- a non-exact differential. Hence, thermodynamic work is rather given by a time-ordered correlation function \cite{PhysRevE.75.050102,RevModPhys.83.771,Hanggi2015}.

To gain more insight into the underlying statistics of quantum work the Two-Time Measurement Formalism \cite{Kurchan2000,Tasaki2000} has proven powerful: In this formulation, a quantum system is prepared in contact with a heat bath of inverse temperature $\beta$. The system is then decoupled from the environment and a projective measurement onto the initial energy eigenbasis is performed. Then, the system is let to evolve before another projective measurement of the energy is performed. As the system is isolated, the work performed on the system is identical to the change in energy. Despite its success, this formalism has several limitations \cite{PhysRevE.93.022131} including the lack of thermodynamic accounting for the measurement process \cite{PhysRevE.94.010103} and its inapplicability to coherently controlled quantum systems \cite{PhysRevLett.118.070601}. Nevertheless, it is important to remark that in complete analogy to how classical mechanics is contained in quantum mechanics (in the appropriate limits)  the Two-Time Measurement Formalism produces work distribution functions which correspond to those of classical systems in semiclassical approximations \cite{Deffner2010,PhysRevX.5.031038, PhysRevE.93.062108, PhysRevE.95.032113, PhysRevE.95.050102}.

To date another decade has gone by, yet quantum stochastic thermodynamics is still rather incomplete. How to describe thermodynamic work and entropy production in open quantum systems is still hotly debated \cite{Deffner2011,Deffner2013EPL,1742-5468-2009-02-P02025, PhysRevB.90.094304, PhysRevLett.102.210401,Leggio2013,Leggio2013a,Campisi2013,Pigeon2016,Santos2017,Strassberg2017}, and with a few exceptions \cite{Deffner2015PRE,Deffner2015PRL,Allahverdyan2016,Gardas2016,Lin2016} most of the literature is restricted to standard Schr\"odinger quantum mechanics. The purpose of the present analysis is to significantly broaden the scope of stochastic thermodynamics, and to take the next, important step -- extend quantum stochastic thermodynamics to quantum field theories. 

In the following, we demonstrate that the Two-Time Measurement Formalism can be systematically used to investigate the work distribution functions of a restricted class of quantum field theories, focusing on a time-dependent version of $\lambda \phi^4$. Closed form expressions for these work distributions are found at leading order, including loop corrections, through the use of a new diagrammatic technique and a mapping between finite-time transition amplitudes and infinite-time scattering amplitudes. It is found that to the perturbative order considered, the work distribution function does not run with the renormalization scale indicating that the distribution is an observable of the quantum field theory. We verify that the quantum Jarzynski and Crooks fluctuation theorems hold exactly and are independent of the renormalization scale. Due to the form of the work distributions, it is straightforward to show that the fluctuation theorems hold if one removes the loop corrections (as would be the case for a classical field theory) and also in the non-relativistic limit.

These results demonstrate that quantum fluctuation theorems and stochastic thermodynamics can be extended to include quantum field theories, our most fundamental theory of nature. Thus, our results open the door for future application of fluctuation theorems to the study of problems at the forefronts of physics -- in condensed matter physics, particle physics, and cosmology.

This paper is organized as follows: In Sec.~\ref{sec:Two-Time-Formalism} we review the Two-Time Measurement Formalism and the quantum Jarzynski equality. We define a restricted class of quantum field theories in Sec.~\ref{sec:Restricting} for which the work distribution function can be calculated. The energy projection operators for a generic real scalar field theory are calculated in Sec.~\ref{sec:Projections-and-Finite-Time} and a method for calculating finite-time transition amplitudes from infinite-time scattering amplitudes is introduced. The mathematical details of this relationship between finite-time and infinite-time amplitudes are detailed in Appendix \ref{app:Finite-to-Infinite}. In Sec.~\ref{sec:Renormalization} we specialize to a time-dependent version of $\lambda \phi^4$ and discuss its renormalization. Then, Sec.~\ref{sec:Generic-Protocol} discusses how closed form expressions for the work distribution function can be calculated at leading order using a graph theoretic technique. The details of the derivation can be found in Appendix \ref{app:Techniques} while the closed form expressions for the work distribution function are in Appendix \ref{app:Work-Distributions}. We discuss the analytic properties of the work distribution function in Sec.~\ref{sec:Distribution-Properties} and analytically verify both the Crooks fluctuation theorem and quantum Jarzynski equality at leading order for time-dependent $\lambda \phi^4$.  In Sec.~\ref{sec:Numerical-Procotols} we numerically evaluate the work distribution function for a relativistic bath and a particular driving protocol, and verify the fluctuation theorems. Interestingly, we find that the dominant process in the work distribution function is particle pair-production through a loop diagram, an effect only found in a quantum field theory. We conclude in Sec.~\ref{sec:Conclusions} with a few remarks.

\section{Preliminaries: Two-Time Measurement Formalism}
\label{sec:Two-Time-Formalism}

We begin by reviewing the Two-Time Measurement Formalism to establish notions and notation \cite{RevModPhys.83.771}: A quantum system is initially, at $t=t_{1}$, in thermal equilibrium with a classical heat bath of inverse temperature $\beta$ \footnote{The choice of an initial Gibbs state is not generic, however it allows one to make contact with classical thermodynamic quantities.}. At $t=t_1+0^{+}$ the system is disconnected from the heat bath and the energy of the system is projectively measured to be $E_{1}$. The system then evolves according to a time dependent protocol until time $t=t_{2}$. At this time, the energy of the system is measured to be $E_{2}$.

Let $\hat{H}(t)$ be the Hamiltonian at time $t$ and let $U(t_{2}, t_{1})$ be the time evolution operator from $t_1$ to $t_2$, and $\hat{\Pi}_{E}$ is the energy projection operator onto the (potentially degenerate) subspace of eigenstates with energy $E$. This projection operator is time-dependent due to the time-dependent Hamiltonian, but for compactness of notation, this dependence will be implicit.

As the system starts in equilibrium, the initial state of the system is given by the thermal density matrix
\be
	\hat{\rho}_{0} = \frac{\e{-\beta \hat{H}(t_1)}}{\tr{\e{-\beta \hat{H}(t_1)}}}.
\label{eq:init-thermal}
\ee
The probability of measuring energy $E_1$ at time $t_1$ is then given by
\be
	P \left( E_1 \right) = \tr{ \hat{\Pi}_{E_{1}} \hat{\rho}_{0}},
\label{eq:prob-E1}
\ee
with the normalized post-measurement state
\be
	\hat{\rho}_{E_{1}} = \frac{\hat{\Pi}_{E_{1}} \hat{\rho}_{0} \hat{\Pi}_{E_{1}} }{\tr{\hat{\Pi}_{E_{1}} \hat{\rho}_{0} \hat{\Pi}_{E_{1}}}}.
\label{eq:post-measurement-E1}
\ee
After being projected into the $E_1$ energy subspace, the system is evolved according to a time-dependent protocol. The conditional probability of measuring energy $E_2$ is
\be
	P \left( E_2 \right| \left. E_1 \right) = \tr{ \hat{\Pi}_{E_{2}} U(t_2, t_1) \hat{\rho}_{E_{1}} U(t_1, t_2)}.
\label{eq:prob-conditional-E2}
\ee
Importantly, the system is isolated from, or at least very weakly coupled to, the heat bath during its evolution. As such, the work performed by the experimenter on the system can be identified with the change in system energy, $W \equiv E_{2} - E_{1}$. One may then define the work distribution function
\be
	\mc{P}(W) = \SumInt_{E_1, E_2} \delta\left( W - E_2 + E_1 \right) P \left( E_1 , E_2 \right).
\label{eq:work-dist-definition}
\ee
Using the definition of the joint probability distribution and Eqs.~\eqref{eq:init-thermal}-\eqref{eq:work-dist-definition},
\begin{equation}
\begin{aligned}
	\mc{P}(W) = & \SumInt_{E_1, E_2} \delta\left( W - E_2 + E_1 \right)\,\frac{ \tr{\hat{\Pi}_{E_{1}}} }{ \tr{ \hat{\Pi}_{E_{1}} \hat{\Pi}_{E_{1}}}}\\
	& \times \tr{\hat{\Pi}_{E_{2}} U(t_2, t_1) \hat{\Pi}_{E_{1}} \hat{\rho}_{0} \hat{\Pi}_{E_{1}} U(t_1, t_2)} .
\end{aligned}
\label{eq:Jarzynski-work-distribution}
\end{equation}
This expression differs from what has been previously shown in the literature due to the presence of the ratio of traces of the projection operators. This is because in previous works the quantum system of interest was assumed to have a discrete energy eigenspectrum. As a consequence, the energy projection operator can be thought of as an idempotent matrix, \textit{i.e.} $\hat{\Pi}_{E_{1}} \hat{\Pi}_{E_{1}} = \hat{\Pi}_{E_{1}}$, and thus this additional term is trivial. However, for systems with a continuum of states the projection operator involves a delta-function which is not idempotent and has non-zero mass dimension. As such, this additional term is essential for proper normalization of the work distribution when one considers a quantum system with a continuum of states.

If the time evolution of system is at least unital \footnote{A unital map is a completely positive map which preserves the identity. More simply, any superposition of unitary quantum maps is a unital map. \cite{Nielsen:2011:QCQ:1972505}}, the quantum Jarzynski equality \cite{Tasaki2000,Kurchan2000,PhysRevE.75.050102,PhysRevA.86.044302,PhysRevE.88.032146, 1742-5468-2013-06-P06016, PhysRevE.89.012127} follows from \eqref{eq:work-dist-definition},
\be
\int dW \, \mc{P} \left( W \right) \e{- \beta W} = \e{- \beta \Delta F}.
\label{eq:quantum-Jarzynski}
\ee
In this expression, $\Delta F$ is the change in free energy from the instantaneous equilibrium distribution at time $t_1$ to time $t_2$.

\section{Restricted Field Theories}
\label{sec:Restricting}

The work distribution function \eqref{eq:Jarzynski-work-distribution} and corresponding quantum Jarzynski equality \eqref{eq:quantum-Jarzynski} are natural objects to consider in the context of non-equilibrium statistical physics. The work distribution function fully classifies all fluctuations involving energy transfer and the quantum Jarzynski equality strongly constrains the form of these fluctuations \cite{Fusco2014}. However, $\mc{P}(W)$, \eqref{eq:Jarzynski-work-distribution}, is not phrased in a natural manner for studying a quantum field theory. The work distribution function requires one to know the energy projection operators, $\hat{\Pi}_{E}$, for the Hamiltonian at the initial and final times. For a generic quantum field theory, the calculation of these operators may prove intractable. Furthermore, \eqref{eq:Jarzynski-work-distribution} is a fundamentally finite-time object as one is performing energy projection measurements at times $t_1$ and $t_2$. Usually, quantum field theory is applied to infinite-time scattering processes as is commonly done in particle physics \cite{Peskin:257493}. This approximation is valid in the context of particle physics because observations are made on timescales significantly greater than the characteristic timescale of particle dynamics. However, non-equilibrium work distributions are of greatest interest when these timescales are comparable.

Given the difficulties associated with the general case, we will restrict the class of quantum field theories and driving protocols which we consider. Working in the rest frame of the experimenter and heat bath, we will assume that the system is governed by a Hamiltonian of the form $H(t) = H_{0} + H_{I}(t)$ where $H_{0}$ is the Hamiltonian for a free field theory. The interacting part of the Hamiltonian is assumed to be sufficiently smooth and have the general form
\be
	H_{I}(t) = \left\{ \begin{array}{lr} H_{I}(t), & \text{for } t \in \left( t_1, t_2 \right) \\
        							0, & \text{otherwise } \end{array} \right. .
\label{eq:Interacting-Hamiltonian}
\ee
It should be noted that these restrictions disallow gauge theories where the matter fields have fixed gauge charges. This is because even in the absence of a classical background field, charged particles self-interact and interact with each other through the exchange of gauge bosons.

Imposing these requirements, it follows that the energy projection operators needed at the beginning and end of the experiment are just those for a free field theory. Furthermore, as will be shown in Sec.~\ref{sec:Projections-and-Finite-Time} and Appendix \ref{app:Finite-to-Infinite}, it will be possible to map the finite-time transition probability onto an infinite-time process because the theory is free at the initial and final times.

These assumptions are essential for our approach in finding the work distribution function. However, we will make an additional set of assumptions for both simplicity and definiteness. For the remainder of this paper, we will restrict ourselves to theories of a single real scalar field, $\phi$, with non-zero mass, $m$. Such theories are described by the Lagrangian
\be
	\mathcal{L} = - \frac{1}{2} \partial_{\mu} \phi \partial^{\mu} \phi - \frac{1}{2} m^2 \phi^2 + \Omega_0 + \mathcal{L}_{\mathrm{int}},
\label{eq:Real-Scalar-Lagrangian}
\ee
where the constant $\Omega_0$ is included to cancel the zero-point energy. Note that we have chosen to work in units where $\hbar=c=1$ and are using the Minkowski metric $\eta_{\mu \nu} = \textrm{diag}\left( -1,+1,+1,+1 \right)$.

Despite their simplicity, such field theories \eqref{eq:Real-Scalar-Lagrangian} have applications across a wide variety of energy scales \cite{Srednicki:1019751, Peskin:257493, Zee:706825}: from phonons \cite{mahan2011condensed, Leutwyler:1996er}, the Ginzburg-Landau theory of superconductivity \cite{Ginzburg:1950sr}, Landau's theory of second order phase transitions \cite{landau1936theory}, and critical phenomena more generally \cite{kleinert2001critical} to the study of spontaneous symmetry breaking \cite{Goldstone1961, PhysRev.127.965}, the Higgs mechanism \cite{PhysRevLett.13.508, PhysRev.145.1156}, and inflationary cosmology \cite{Baumann:2009ds}.

\begin{widetext}
\section{Projection Operators And Finite-Time Transitions}
\label{sec:Projections-and-Finite-Time}

Given the form of the interaction \eqref{eq:Interacting-Hamiltonian}, the energy projection operators are the free theory projection operators. Note that the free Hamiltonian commutes with the number operator. Hence, we can express energy projection operators as a sum over projections with definite energy, $E$, and particle number $n$. They can be written as
\begin{equation}
\hat{\Pi}_{E,n} =  \int \widetilde{d^{3}k_{1}} \, \ldots \, \widetilde{d^{3}k_{n}} \;  \delta \left( E - \omega_1 - \ldots - \omega_n \right) \frac{1}{n!} \left| k_1, \ldots, k_n \right\rangle \left \langle k_1, \ldots, k_n \right| ,
\end{equation}
where $\omega_j = (m^2 + k_{j}^{2})^{(1/2)}$ is the energy of the $j$th particle and $\widetilde{d^{3}k_{j}} = d^{3} k_{j} / \left( 2 \pi \right)^3 2 \omega_j$ is the Lorentz invariant measure \cite{Srednicki:1019751}. Summing over energetically degenerate subspaces we can further write $\hat{\Pi}_{E} = \sum_n \hat{\Pi}_{E,n}$. Even though the field theory has a mass gap, this general form holds for all energy projection operators, including the ground state projection with $E=0$.

Returning to the work distribution function \eqref{eq:Jarzynski-work-distribution} and making use of these definitions for the energy projection operators, we obtain
\begin{equation}
\label{eq:Work-Dist-Scattering}
\mc{P}(W) = \sum_{n_1,n_2}\int \prod_{i}^{n_1} \prod_{j}^{n_2} \widetilde{d^{3}k_{i}} \, \widetilde{d^{3}k_{j}'}\, \delta \left( W + \sum_{l=1}^{n_1} \omega_l - \sum_{l=1}^{n_2} \omega_{l}' \right) \left| \frac{ \left \langle k_1 ', \ldots, k_{n_2}' \right| U(t_2, t_1) \left| k_1, \ldots, k_{n_1} \right\rangle }{ \sqrt{n_2 ! n_1 !} } \right|^2 \frac{\e{-\beta \sum_{l=1}^{n_1} \omega_l }}{\tr{\e{-\beta \hat{H}_{0}}}} .
\end{equation}
The distribution \eqref{eq:Work-Dist-Scattering} is normalized by the free energy of the free field theory, $\mrm{tr}\{ \exp{(-\beta \hat{H}_{0})}\}= \exp{(-\beta F_{0})}$. The momenta of the incoming and outgoing particles are integrated over in a Lorentz invariant manner and thus the integration measure is frame independent. Furthermore, each incoming particle is associated with a Boltzmann weight $\e{-\beta \omega}$. The single delta-function ensures conservation of energy. Lastly, the quantity $\left \langle k_1 ', \ldots, k_{n_2}' \right| U(t_2, t_1) \left| k_1, \ldots, k_{n_1} \right\rangle$ is the finite-time transition amplitude for the time-dependent system.

To make use of the machinery of quantum field theory, it will be necessary to rewrite this finite-time amplitude in terms of an infinite-time scattering process. The mathematical details are in Appendix \ref{app:Finite-to-Infinite}, but a high-level description and the intuition for the mapping are provided here. 

Due to the restrictions placed on the form of the interaction Hamiltonian \eqref{eq:Interacting-Hamiltonian}, the quantum field theory is free at the initial and final times. One can imagine extending the finite-time experiment outside of the interval $\left[t_{1}, t_{2} \right]$ by assuming the Hamiltonian remains non-interacting before and after the projective energy measurements. As the Hamiltonian is time independent for $t\leq t_1$ and $t\geq t_2$, no additional work is performed and the work distribution function is identical to the finite-time process. Furthermore, as the projective measurements place the system in an energy eigenstate of the free theory at the initial and final times, these states can be evolved arbitrarily far into the past or future, respectively, in the Schr\"odinger picture at the cost of an overall, yet irrelevant, phase. Thus, we map the finite-time transition amplitude onto an infinite-time scatttering process, and we find
\begin{equation}
\begin{split}
&\left| \left \langle k_1 ', \ldots, k_{n_2}' \right| U(t_2, t_1) \left| k_1, \ldots, k_{n_1} \right\rangle \right| = \\
&\quad \left| \int \right. d^{3}x_{1}' d^{3}x_{1} \ldots \; \e{-i k_{1}' x_{1}'} \e{i k_{1} x_{1}}\ldots\doubleoverarrow{\partial_{0_{x_{1}'}}} \doubleoverarrow{\partial_{0_{x_{1}}}} \ldots \left. \cdot {}_{I}\left\langle \Omega \right| T \left[ U_{I}(\infty,-\infty) \phi_{I}(x_1 ') \ldots \phi_{I}(x_1) \ldots \right] \left| \Omega \right\rangle_{I} \vphantom{\int} \right|.
\end{split}
\label{eq:Finite-Time-To-QFT-mainbody}
\end{equation}
In this expression, the subscript $I$ is used to indicate operators in the Interaction picture. The state $\left| \Omega \right\rangle_{I}$ is defined as the vacuum state of the free theory, \textit{i.e.} $\hat{H}_0 \left| \Omega \right\rangle_{I} = 0$. We also have $f \doubleoverarrow{\partial_{\mu}} g \equiv f \left( \partial_{\mu} g \right) - \left( \partial_{\mu} f \right) g$, see Ref. \cite{Srednicki:1019751}.
\end{widetext}

\section{Renormalization of time-dependent theories}
\label{sec:Renormalization}

For non-trivial work to be performed on the system, the interaction Hamiltonian \eqref{eq:Interacting-Hamiltonian} must be time-dependent. This time-dependence breaks Lorentz invariance by singling out a preferred frame, the experimenter's frame. Thus, quantities such as energy and time are always measured with respect to this frame. This differs significantly from the usual approach to quantum field theory where Lorentz invariance essential \cite{Peskin:257493}. As such, significant care must be taken in the definition and renormalization of the quantum field theory.

\paragraph*{Formulation}

Generally, we may choose any time-dependent interaction in \eqref{eq:Real-Scalar-Lagrangian}, however we will focus on a time-dependent variant of $\lambda \phi^4$, and we have,
\be
	\mathcal{L}_{\textrm{int}} = - \frac{1}{4!} \lambda(t) \phi^4.
\label{eq:Time-dependent-Interaction-Lagrangian}
\ee
The time-independent $\lambda \phi^4$ is a renormalizable field theory \cite{Srednicki:1019751, Peskin:257493, Zee:706825}, which can be shown rigorously through Dyson-Weinberg power counting arguments \cite{PhysRev.75.1736, PhysRev.118.838}. Being renormalizable, the theory only requires a finite number of counterterms to cancel divergences due to loop corrections and is valid at all energy scales, up to considerations of strong coupling. However, these power counting arguments rely on the Lorentz invariance of the field theory's Lagrangian density. As Lorentz invariance is broken in \eqref{eq:Time-dependent-Interaction-Lagrangian}, it is not clear that this theory can be renormalized with a finite number of counterterms. 

A mathematically equivalent, but more intuitive approach, is to rewrite this field theory as a non-renormalizable effective field theory with a classical source. This is done by promoting $\lambda$ to a classical, non-dynamical, scalar field $\chi_{\mathrm{cl}}$ with mass $M$. This mass scale is assumed to be much greater than any other energy scale in the system and sets the cut-off scale for this effective field theory. As a book-keeping mechanism, it will be convenient to introduce a dimensionless parameter $g=1$ to keep track of the perturbative expansion as the theory no longer has an explicit coupling constant. The Lagrangian density then becomes
\be
	\mathcal{L} = - \frac{1}{2} \partial_{\mu} \phi \partial^{\mu} \phi - \frac{1}{2} m^2 \phi^2 + \Omega_0  - \frac{g}{4! M} \chi_{\mathrm{cl}} \phi^4.
\label{eq:Time-dependent-EFT}
\ee
Additional interaction terms induced by the breaking of Lorentz invariance are suppressed by increasing powers of $\frac{g}{M}$. Equation~\eqref{eq:Time-dependent-EFT} can be thought of as the leading order expression of \eqref{eq:Time-dependent-Interaction-Lagrangian} as an effective field theory. The interaction term in this theory has mass-dimension five and thus this theory is non-renormalizable \cite{Peskin:257493}. Being non-renormalizable, an infinite set of counterterms is required to cancel divergences and the theory may only be applied at energy scales up to its cutoff, $M$. For present purposes, the counterterms of interest may be expressed as
\be
	\mathcal{L}_{\mathrm{ctr}} = - \sum_{j,k} c_{j,k} \frac{g^j}{M^j} \chi_{\mathrm{cl}}^j \phi^k .
\label{eq:Counterterm-Lagrangian}
\ee

One key advantage of the effective field theory  \eqref{eq:Time-dependent-EFT} over Eq.~\eqref{eq:Time-dependent-Interaction-Lagrangian} is that the classical field $\chi_{\mathrm{cl}}$ can be thought of as a work reservoir \cite{Callen1985,Deffner2013}. This reservoir sources all interactions and the $\chi_{\mathrm{cl}}$ field carries this energy into or out of the system. In the present case, this leads to more intuitive Feynman diagrams where energy is conserved at every vertex as opposed to the theory described in Eq.~\eqref{eq:Time-dependent-Interaction-Lagrangian} where vertices only include $\phi$, and hence do not conserve energy. Note, however, that the two approaches are mathematically fully equivalent and we may freely switch between them by identifying $g/M\,\chi_{\mathrm{cl}}(t) = \lambda(t)$.

\paragraph*{Renormalization}
\begin{figure*}[t]
	\hspace*{\fill}
	\subfloat[Loop correction to the propagator. \label{fig:two-point-loop}]{
	\begin{fmffile}{fgraphs01}
		\begin{fmfgraph*}(60,60)
		\fmfset{thick}{1.25}
		\fmfpen{thick}
		\fmfforce{(.05w,.3h)}{i1} 
		\fmfforce{(.2w,.3h)}{i11} 
		\fmfforce{(.95w,.3h)}{o1}
		\fmfforce{(.8w,.3h)}{o11}
		\fmfforce{(.5w,.75h)}{t1}
		\fmfforce{(.5w,.05h)}{b1}
		\fmfforce{(.5w,.6h)}{t11}
		\fmf{plain, tension=100}{i1,i11}
		\fmf{plain, tension=100}{i11,v1}
		\fmf{plain, tension=100}{v1,o11}
		\fmf{plain, tension=100}{o11,o1}
		\fmf{plain, left, tension=.1}{v1,b1,v1}
		\fmf{dashes, tension = .1, label=\large$\chi_{\mathrm{cl}}$, label.dist=7.5,label.side=left}{t1,t11}
		\fmf{dashes, tension = .1}{t11,v1}
		\fmfdot{v1}
		\end{fmfgraph*}
	\end{fmffile}
	} \hfill
	\subfloat[Counterterm for the propagator. \label{fig:two-point-counterterm}]{
	\begin{fmffile}{fgraphs02} 
		\begin{fmfgraph*}(60,60)
		\fmfset{thick}{1.25}
		\fmfpen{thick}
		\fmfforce{(.05w,.3h)}{i1} 
		\fmfforce{(.2w,.3h)}{i11} 
		\fmfforce{(.95w,.3h)}{o1}
		\fmfforce{(.8w,.3h)}{o11}
		\fmfforce{(.5w,.75h)}{t1}
		\fmfforce{(.5w,.05h)}{b1}
		\fmfforce{(.5w,.6h)}{t11}
		\fmf{plain, tension=100}{i1,i11}
		\fmf{plain, tension=100}{i11,v1}
		\fmf{plain, tension=100}{v1,o11}
		\fmf{plain, tension=100}{o11,o1}
		\fmf{dashes, tension = .1, label=\large$\chi_{\mathrm{cl}}$, label.dist=7.5,label.side=left}{t1,t11}
		\fmf{dashes, tension = .1}{t11,v1}
		\fmfv{decoration.shape=cross, decoration.size=10}{v1}
		\end{fmfgraph*}
	\end{fmffile}
	}
	\hspace*{\fill} \newline
	\caption{Leading order corrections to the propagator of the scalar field $\phi$. The interactions are sourced by insertions of the classical, non-dynamical field $\chi_{\mathrm{cl}}$. \label{fig:two-point-diagrams}}
\end{figure*}
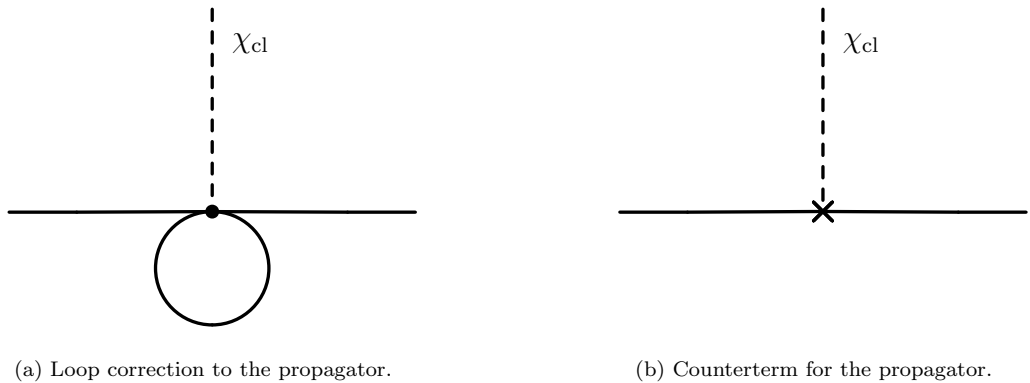
\begin{figure*}[t]
	\hspace*{\fill}
	\subfloat[Two loop vacuum diagram. \label{fig:vacuum-loop}]{
	\begin{fmffile}{fgraphs03} 
		\begin{fmfgraph*}(45,45)
		\fmfset{thick}{1.25}
		\fmfpen{thick}
		\fmfforce{(0,.25h)}{i1} 
		\fmfforce{(w,.25h)}{o1}
		\fmfforce{(.5w,.8h)}{t1}
		\fmfforce{(.5w,.6h)}{t11}
		\fmf{phantom, tension=100}{i1,p11,p12,v1}
		\fmf{phantom, tension=100}{v1,p21,p22,o1}
		\fmf{dashes, tension = .1, label=\large$\chi_{\mathrm{cl}}$, label.dist=7.5,label.side=left}{t1,t11}
		\fmf{dashes, tension = .1}{t11,v1}
		\fmf{plain, right}{v1,p11,v1}
		\fmf{plain, right}{v1,p22,v1}
		\fmfdot{v1}
		\end{fmfgraph*}
	\end{fmffile}
	} \hfill
	\subfloat[Counterterm loop vacuum diagram. \label{fig:vacuum-counterterm-loop}]{
	\begin{fmffile}{fgraphs04} 
		\begin{fmfgraph*}(45,45)
		\fmfset{thick}{1.25}
		\fmfpen{thick}
		\fmfforce{(0,.25h)}{i1} 
		\fmfforce{(w,.25h)}{o1}
		\fmfforce{(.5w,.8h)}{t1}
		\fmfforce{(.5w,.6h)}{t11}
		\fmf{phantom, tension=100}{i1,p11,p12,v1}
		\fmf{phantom, tension=100}{v1,p21,p22,o1}
		\fmf{dashes, tension = .1, label=\large$\chi_{\mathrm{cl}}$, label.dist=7.5,label.side=left}{t1,t11}
		\fmf{dashes, tension = .1}{t11,v1}
		\fmf{phantom, right}{v1,p11,v1}
		\fmf{plain, right}{v1,p22,v1}
		\fmfv{decoration.shape=cross, decoration.size=10}{v1}
		\end{fmfgraph*}
	\end{fmffile}
	} \hfill
	\subfloat[Counterterm vacuum diagram. \label{fig:vacuum-counterterm}]{
	\begin{fmffile}{fgraphs05} 
		\begin{fmfgraph*}(45,45)
		\fmfset{thick}{1.25}
		\fmfpen{thick}
		\fmfforce{(0,.25h)}{i1} 
		\fmfforce{(w,.25h)}{o1}
		\fmfforce{(.5w,.8h)}{t1}
		\fmfforce{(.5w,.6h)}{t11}
		\fmf{phantom, tension=100}{i1,v1,o1}
		\fmf{dashes, tension = .1, label=\large$\chi_{\mathrm{cl}}$, label.dist=7.5,label.side=left}{t1,t11}
		\fmf{dashes, tension = .1}{t11,v1}
		\fmfv{decoration.shape=cross, decoration.size=10}{v1}
		\end{fmfgraph*}
	\end{fmffile}
	}
	\hspace*{\fill} \newline
	\caption{Vacuum energy contributions of $\chi_{\mathrm{cl}}$. The interactions are sourced by insertions of the classical, nondynamical field $\chi_{\mathrm{cl}}$. \label{fig:vacuum-diagrams}}
\end{figure*}
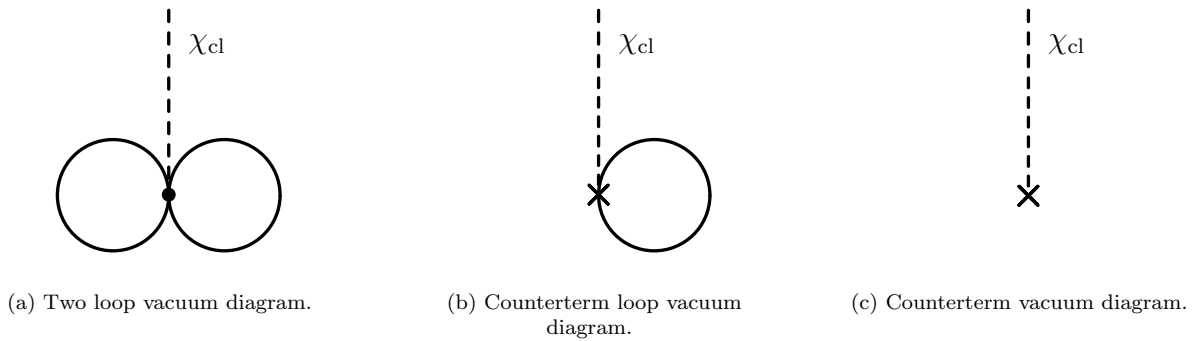
We will be working to leading order in the perturbative parameter $g=1$. However, even at this order, we must consider loop diagrams which are formally divergent. We will use dimensional regularization in $d = 4 - \epsilon$ dimensions to parameterize the divergences and work within the framework of the $\overline{MS}$ renormalization scheme at an energy scale $\mu$ to systematically assign values to the counterterms, see \textit{e.g.} Refs. \cite{Srednicki:1019751, Peskin:257493, Zee:706825}. At leading order, there are only two divergent diagrams we will need to consider. The first is the loop correction to the $\phi$ propagator, Fig.~\ref{fig:two-point-diagrams}. The second is the vacuum energy diagram sourced by the classical field $\chi_{\mathrm{cl}}$, Fig.~\ref{fig:vacuum-diagrams}. In these diagrams, the scalar field $\phi$ is denoted by a solid line while the classical background field $\chi_{\mathrm{cl}}$ is represented by a dotted line.

We first consider the corrections to the $\phi$ propagator shown in Fig.~\ref{fig:two-point-diagrams}. At leading order, the propagator is modified by a loop correction, Fig.~\ref{fig:two-point-loop}, whose divergent part is canceled by a counterterm, Fig.~\ref{fig:two-point-counterterm}. The relevant counterterm from the Lagrangian \eqref{eq:Counterterm-Lagrangian} is $c_{1,2} \frac{g}{M} \chi_{\mathrm{cl}} \phi^2$. Working in $d=4-\epsilon$ dimensions, the $\overline{MS}$ renormalization scheme requires us to fix
\be
	c_{1,2} = \frac{1}{2} \left( \frac{m}{4 \pi} \right)^2 \frac{1}{\epsilon}.
\ee
In the limit of $\chi_{\mathrm{cl}}$ being a time independent background, the remaining finite part of the loop-diagram matches with that of regular $\lambda \phi^4$ theory.

The diagrams in Fig.~\ref{fig:vacuum-diagrams} are used to calculate the change in vacuum energy of the $\phi$ field due to the classical background field $\chi_{\mathrm{cl}}$. The two loop diagram in Fig.~\ref{fig:vacuum-loop} is the vacuum bubble induced by the $\chi_{\mathrm{cl}}$ background. Figure~\ref{fig:vacuum-counterterm-loop} includes the contribution of the counterterm fixed by the loop corrections to the propagator while Fig.~\ref{fig:vacuum-counterterm} corresponds to the contribution of the $c_{1,0} \frac{g}{M} \chi_{\mathrm{cl}}$ counterterm. As the counterterm in Fig.~\ref{fig:vacuum-counterterm-loop} is already fixed by the propagator, the $\overline{MS}$ scheme requires the choice
\be
	c_{1,0} = \frac{1}{2} \left( \frac{m}{4 \pi} \right)^4 \frac{1}{\epsilon^2}.
\ee
The remaining finite part of the diagrams in Fig.~\ref{fig:vacuum-diagrams} is given by
\be
	\ldots = - \frac{i}{8} \left( \frac{m}{4 \pi} \right)^4 \int d^{4}z \, \frac{g}{M} \chi_{\mathrm{cl}}(z).
\label{eq:incomplete-vacuum-energy}
\ee
The expression \eqref{eq:incomplete-vacuum-energy} involves the integral over all space and time of the background field $\chi_{\mathrm{cl}}$. This quantity will be formally infinite unless the system is restricted to a large but finite spatial volume $V$. As $\chi_{\mathrm{cl}}$ is spatially uniform in the experimenter's frame, it then follows that
\be
	\ldots = -\frac{i}{8} \left( \frac{m}{4 \pi} \right)^4 V \int dt \, \frac{g}{M} \chi_{\mathrm{cl}}(t).
\label{eq:vacuum-energy}
\ee

\paragraph*{Disconnected Vacuum Diagrams}
In the standard framework of quantum field theory, one assumes that all interactions are switched on and off adiabatically in the distant past and future. Consequently, for a field theory with a mass gap, one can use the adiabatic theorem to show that the disconnected vacuum diagrams only contribute an irrelevant overall phase to any scattering amplitude \cite{nenciu1989adiabatic, doi:10.1063/1.2740469}. However, non-equilibrium evolution requires that the interaction parameters are varied non-adiabatically. As such, one must include the disconnected vacuum diagrams in the calculation of scattering amplitudes.

Consider an $n$-point correlation function of the form $\left\langle \Omega \right| T \left[ U_{I}(\infty,-\infty) \phi_{I}(x_{1}) \ldots \right] \left| \Omega \right\rangle$. For a contributing Feynman diagram, we will call any part of the diagram that can be traced to an external field source $\phi_{I}(x_{i})$ connected. The contributions of all such connected diagrams will be denoted by $\left\langle \Omega \right| T \left[ U_{I}(\infty,-\infty) \phi_{I}(x_{1}) \ldots \right] \left| \Omega \right\rangle_{C}$. Any other component of the diagram will be considered a disconnected vacuum subdiagram. Note that all such vacuum diagrams necessarily involve the background field $\chi_{\mathrm{cl}}$ as it sources all interactions. The contribution of the set of disconnected vacuum diagrams is given by $\left\langle \Omega \right| U_{I}(\infty,-\infty) \left| \Omega \right\rangle$.

The $n$-point correlation function factorizes into the product of the connected $n$-point  diagrams, $\left\langle \Omega \right| T \left[ U_{I}(\infty,-\infty) \phi_{I}(x_{1}) \ldots \right] \left| \Omega \right\rangle_{C}$, and the vacuum diagrams, $\left\langle \Omega \right| U_{I}(\infty,-\infty) \left| \Omega \right\rangle$. The vacuum diagram contribution $\left\langle \Omega \right| U_{I}(\infty,-\infty) \left| \Omega \right\rangle$ has the property that it can be expressed as the exponential of the sum of all unique vacuum diagrams. This is due to the fact that if multiple copies of the same vacuum subdiagram are present in a Feynman diagram, one must divide by the number of possible rearrangements of these identical diagrams. Thus,
\begin{equation}
\begin{split}
	&\left\langle \Omega \right| T \left[ U_{I}(\infty,-\infty) \phi_{I}(x) \ldots \right] \left| \Omega \right\rangle \\
	&\quad\quad= \begin{aligned}[t] &\e{\sum \textrm{Vacuum Diagrams}}  \\
	& \times\la \Omega \right| T \left[ U_{I}(\infty,-\infty) \phi_{I}(x) \ldots \right] \left| \Omega \ra_{\mathrm{C}} . \end{aligned}
\end{split}
\end{equation}
As was shown in Eq.~\eqref{eq:vacuum-energy}, the leading order vacuum diagram is purely imaginary. Therefore, the disconnected vacuum diagrams only contribute an overall phase at leading order, and thus one only needs to consider diagrams connected to the field sources.

From a thermodynamic perspective, the failure of disconnected vacuum diagrams to contribute to the work distribution function is expected. Disconnected vacuum diagrams by definition cannot involve field sources of $\phi$ and thus cannot involve the transfer of energy into or out of the system.

\section{Work in Quantum Field Theories}
\label{sec:Generic-Protocol}

\paragraph*{Trivial and Non-Trivial Scattering}
As seen in Sec.~\ref{sec:Projections-and-Finite-Time} with details provided in Appendix \ref{app:Finite-to-Infinite}, the finite-time transition probability can be calculated from an infinite-time scattering process. From \eqref{eq:Finite-Time-To-QFT-mainbody} it can be shown that this requires the evaluation of an $n$-point correlation function in the Interaction picture. This can naturally be done by perturbatively expanding the time evolution operator in terms of the Dyson series and subsequently applying Wick's theorem to evaluate the resulting free-field correlation functions. At leading order in perturbation theory we have
\begin{equation}
\begin{split}
	U_{I}\left( \infty, - \infty \right) & = \mc{T}_> \left[ \e{-i \int_{-\infty}^{\infty} H_{\mathrm{int}}(t) \, dt} \right] \\
	& \approx \eins - i \int_{-\infty}^{\infty} d t \; H_{\mathrm{int}}(t) .
\label{eq:Dyson-Series}
\end{split}
\end{equation}
In this expression $H_{\mathrm{int}}(t)$ is the interaction Hamiltonian in the Interaction picture. Explicitly, it is given by
\begin{equation}
\begin{aligned}
H_{\mathrm{int}} = \int & d^{3}x \left[ \vphantom{\sum} \right. \frac{g}{4! M} \chi_{\mathrm{cl}} \left( x \right) \phi_{I}^4\left( x \right) \\
 & \left. + c_{1,0} \frac{g}{M} \chi_{\mathrm{cl}}\left( x \right) + c_{1,2} \frac{g}{M} \chi_{\mathrm{cl}} \left( x \right) \phi_{I}^2\left( x \right) \right].
\end{aligned}
\label{eq:Interaction-Hamiltonian}
\end{equation}
Using Eq.~\eqref{eq:Dyson-Series} in the scattering amplitude \eqref{eq:Finite-Time-To-QFT-mainbody} schematically yields an expression of the form $\left| \left\langle \textrm{out} \right| \left. \textrm{in} \right\rangle \right|^2 + \left| \left\langle \textrm{out} \right| H \left| \textrm{in} \right\rangle \right|^2$. This is the sum of two terms with distinct physical origins: The first term, $\left| \left\langle \textrm{out} \right| \left. \textrm{in} \right\rangle \right|^2$, is the scattering amplitude for the trivial process where the perturbation does not enter and no work is performed on the system. As no work is performed, this will contribute a delta-function to the work distribution. The second term, $\left| \left\langle \textrm{out} \right| H \left| \textrm{in} \right\rangle \right|^2$, involves non-trivial scattering through the time-dependent perturbation. The combined probability distributions of these two processes, however, will not integrate to unity. This is a consequence of working at finite order in the Dyson series, \eqref{eq:Dyson-Series}. The approximation violates unitarity, which generally has to be imposed by hand, see for instance Ref.~\cite{Acconcia2017}.

In the present case, however, it is possible to sidestep this issue since the Jarzynski equality holds separately for the non-trivial component of the work distribution, $\rho(W)$. The full work distribution has the general form $\mc{P}(W) = a \, \delta(W) + \rho(W)$ where $a = 1 - \int dW \, \rho(W)$ is a positive constant chosen to impose unitarity. Given the restrictions on the interaction Hamiltonian imposed by \eqref{eq:Interacting-Hamiltonian}, the system starts and ends as a free field theory and thus $\Delta F = 0$. From the Jarzynski equality, $1 = \int dW \, \mc{P} \left( W \right) \e{- \beta W}$, we can write,
\begin{equation}
\begin{split}
1 - a & = \int dW \, \rho(W) \, \e{-\beta W} \\
\Rightarrow \int dW \, \rho(W) & = \int dW \, \rho(W) \, \e{-\beta W}
\end{split}
\end{equation}
In conclusion, the Jarzynski equality \eqref{eq01} holds for the normalized, non-trivial part of the work distribution. Therefore, as the trivial component of the scattering process only contributes a delta-function to the work distribution and does not impact the Jarzynski equality, it suffices to consider the non-trivial part of $\mc{P}(W)$.

\paragraph*{Calculational Approach}
Several complications arise in the treatment of the non-trivial scattering term. The interaction Hamiltonian \eqref{eq:Interaction-Hamiltonian} is composed of terms which involve the scattering of at most four incoming or outgoing particles. As the general work distribution function, \eqref{eq:Work-Dist-Scattering}, involves any number of incoming or outgoing particles, the Feynman diagrams which describe these processes will be composed of several disconnected subdiagrams. One must sum over all possible permutations of these subdiagrams before squaring the resulting amplitude. This is in stark contrast to the usual procedure in quantum field theory where one is only interested in fully connected diagrams and their permutations. To further complicate the matter, even once one has the square of the amplitude of all permutations, one still must integrate over all momenta and sum over all possible particle numbers as proscribed in \eqref{eq:Work-Dist-Scattering}. Carrying out this procedure in generality proves a formidable challenge to a direct application of existing field theoretic techniques.

In this work, we instead pioneer a graph theoretic approach which allows us to classify the products of Feynman diagrams in such a manner that the infinite sums over particle number can be carried out exactly. This leads to closed form expressions for the leading order work distribution where only a few kinematic integrals must be performed. The details of this procedure are in Appendix~\ref{app:Techniques} but a brief description is provided here. While $\left| \left\langle \textrm{out} \right| H \left| \textrm{in} \right\rangle \right|^2$ can be thought of as the square of the sum of all permutated diagrams, it will be more helpful to think of it in terms of the sum over the crossterms of two permutated diagrams. The incoming and outgoing field sources of each diagram are labeled by integers up to $n_1$ and $n_2$ respectively. One proceeds to ``glue'' the two diagrams together by identifying the corresponding field sources in each diagram. The resulting ``glued'' diagram can then be classified in terms of its graph topology, specifically the topology of the connected subgraph(s) which contain insertions of the interaction Hamiltonian. Rephrased in this language, the combinatorics of the sum over permutations and subsequent sum over particle number becomes tractable.

Ultimately, one finds that the work distribution function is naturally written as the sum of five distributions: the work distribution for when the particle number is unchanged, the distributions for when the particle number increases or decreases by two, and the distributions for when the particle number increases or decreases by four. These are denoted by the distributions $\rho_{n \rightarrow n}(W)$, $\rho_{n \rightarrow n \pm 2}(W)$, and $\rho_{n \rightarrow n \pm 4}(W)$ respectively. It should be stressed that the subscript $n$ in these distributions does not correspond to a specific particle number as the particle number has been summed over. Closed form, unnormalized, expressions for these five distributions are given in Appendix~\ref{app:Work-Distributions}.

\section{Analytic Properties}
\label{sec:Distribution-Properties}

\paragraph*{Form of the Work Distributions}
As an example for the five contributions to $\mc{P}(W)$, we discuss $\rho_{n \rightarrow n + 2}(W)$ in detail as it illustrates all key properties. This distribution may be decomposed into two components,
\be
\rho_{n \rightarrow n + 2} \left( W \right) = \rho^{\textrm{tree}}_{n \rightarrow n + 2} \left( W \right) + \rho^{\textrm{loop}}_{n \rightarrow n + 2} \left( W \right) ,
\ee
where $\rho^{\textrm{tree}}_{n \rightarrow n + 2} \left( W \right)$ is the distribution of work arising from tree-level processes and $\rho^{\textrm{loop}}_{n \rightarrow n + 2} \left( W \right)$ originates from diagrams involving a loop. In a loose sense, $\rho^{\textrm{tree}}_{n \rightarrow n + 2} \left( W \right)$ can be thought of as ``classical'' contributions to the work distribution as tree-level diagrams satisfy the classical equations of motion. The distribution $\rho^{\textrm{loop}}_{n \rightarrow n + 2} \left( W \right)$ corresponds to processes which violate the classical equations of motion and are purely a result of second quantization. Only the distributions $\rho_{n \rightarrow n + 2}(W)$ and $\rho_{n \rightarrow n - 2}(W)$ have contributions from loop diagrams at this order.
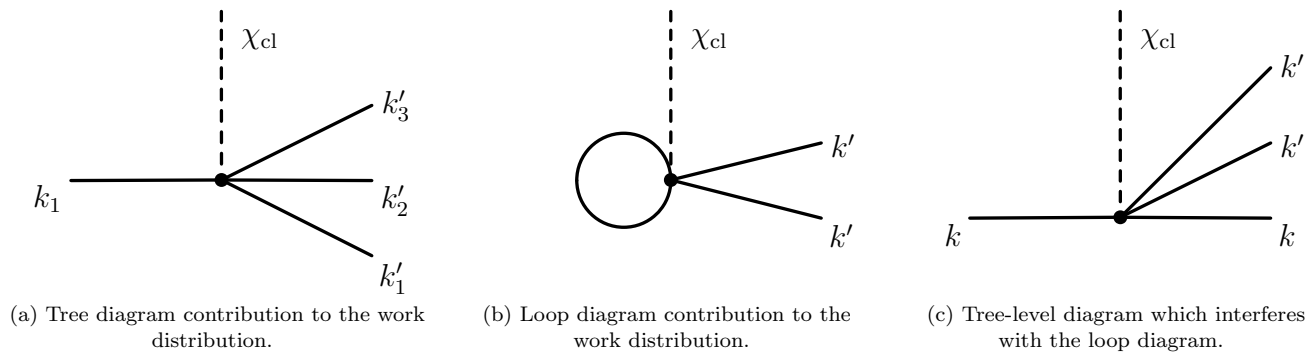
\begin{figure*}[t]
	\hspace*{\fill}
	\subfloat[Tree diagram contribution to the work distribution. \label{fig:tree-n-to-nplus2}]{
	\begin{fmffile}{fgraphs06}
		\begin{fmfgraph*}(50,50)
		\fmfset{thick}{1.25}
		\fmfpen{thick}
		\fmfforce{(.1w,.3h)}{i1} 
		\fmfforce{(.9w,.1h)}{o1}
		\fmfforce{(.9w,.3h)}{o2}
		\fmfforce{(.9w,.5h)}{o3}
		\fmfforce{(.5w,.75h)}{t1}
		\fmfforce{(.5w,.6h)}{t11}
		\fmf{plain, tension=100}{i1,v1}
		\fmf{plain, tension=.1}{v1,o1}
		\fmf{plain, tension=100}{v1,o2}
		\fmf{plain, tension=.1}{v1,o3}
		\fmf{dashes, tension = .1, label=\large$\chi_{\mathrm{cl}}$, label.dist=7.5,label.side=left}{t1,t11}
		\fmf{dashes, tension = .1}{t11,v1}
		\fmfdot{v1}
		\fmflabel{\large$k_1$}{i1}
		\fmflabel{\large$k_{1}'$}{o1}
		\fmflabel{\large$k_{2}'$}{o2}
		\fmflabel{\large$k_{3}'$}{o3}
		\end{fmfgraph*}
	\end{fmffile}
	} 
	\hspace*{\fill}
		\subfloat[Loop diagram contribution to the work distribution. \label{fig:loop-n-to-nplus2}]{
	\begin{fmffile}{fgraphs07}
		\begin{fmfgraph*}(50,50)
		\fmfset{thick}{1.25}
		\fmfpen{thick}
		\fmfforce{(.25w,.3h)}{b1}
		\fmfforce{(.1w,.3h)}{i1} 
		\fmfforce{(.2w,.3h)}{i11} 
		\fmfforce{(.9w,.3h)}{o1}
		\fmfforce{(.9w,.2h)}{o2}
		\fmfforce{(.9w,.4h)}{o3}
		\fmfforce{(.5w,.75h)}{t1}
		\fmfforce{(.5w,.6h)}{t11}
		\fmf{phantom, tension=100}{i1,v1}
		\fmf{phantom, tension=100}{v1,o1}
		\fmf{plain, tension=.1}{v1,o2}
		\fmf{plain, tension=.1}{v1,o3}
		\fmf{plain, left, tension=.1}{v1,b1,v1}
		\fmf{dashes, tension = .1, label=\large$\chi_{\mathrm{cl}}$, label.dist=7.5,label.side=left}{t1,t11}
		\fmf{dashes, tension = .1}{t11,v1}
		\fmfdot{v1}
		\fmflabel{\large$k'$}{o2}
		\fmflabel{\large$k'$}{o3}
		\end{fmfgraph*}
	\end{fmffile}
	} 
	\hspace*{\fill}
	\subfloat[Tree-level diagram which interferes with the loop diagram. \label{fig:loop-tree-n-to-nplus2}]{
	\begin{fmffile}{fgraphs08}
		\begin{fmfgraph*}(50,50)
		\fmfset{thick}{1.25}
		\fmfpen{thick}
		\fmfforce{(.1w,.2h)}{i1} 
		\fmfforce{(.9w,.2h)}{o1}
		\fmfforce{(.9w,.4h)}{o2}
		\fmfforce{(.9w,.6h)}{o3}
		\fmfforce{(.5w,.75h)}{t1}
		\fmfforce{(.5w,.6h)}{t11}
		\fmf{plain, tension=100}{i1,v1}
		\fmf{plain, tension=100}{v1,o1}
		\fmf{plain, tension=.1}{v1,o2}
		\fmf{plain, tension=.1}{v1,o3}
		\fmf{dashes, tension = .1, label=\large$\chi_{\mathrm{cl}}$, label.dist=7.5,label.side=left}{t1,t11}
		\fmf{dashes, tension = .1}{t11,v1}
		\fmfdot{v1}
		\fmflabel{\large$k$}{i1}
		\fmflabel{\large$k$}{o1}
		\fmflabel{\large$k'$}{o2}
		\fmflabel{\large$k'$}{o3}
		\end{fmfgraph*}
	\end{fmffile}
	}
	\newline
	\caption{Diagrams which contribute to the work distribution function $\rho_{n \rightarrow n + 2}(W)$. \label{fig:work-diagrams-n-to-nplus2}}
\end{figure*}

We begin with the tree-level contribution. While both the incoming and outgoing states will potentially involve many particles, the relevant subdiagram generated by the interaction Hamiltonian is shown in Fig.~\ref{fig:tree-n-to-nplus2}. This is simply the tree-level process where one particle becomes three. The distribution of resulting work done on the system is given by \footnote{For convenience, we have chosen to work in terms of $\lambda(t)$ rather than the mathematically equivalent $g/M \, \chi_{\mathrm{cl}}\left( t \right)$.},
\begin{widetext}
\begin{equation}
\begin{split}
&\rho^{\textrm{tree}}_{n \rightarrow n + 2} \left( W \right) =  \frac{V}{3!} \, \left| \int dt \, \lambda(t) \e{i W t} \right|^2 \int \widetilde{d^{3}k_{1}} \, \widetilde{d^{3}k_{1}'} \, \widetilde{d^{3}k_{2}'} \, \widetilde{d^{3}k_{3}'} \; \delta \left( W + \omega_{1} - \omega_{1}' - \omega_{2}' -\omega_{3}' \right) \\
& \quad\times \left( 2 \pi \right)^3 \delta^3 \left( k_1 - k_1' - k_2' - k_3' \right) \; \left( \frac{1}{\e{\beta \omega_1} - 1} \right) \left( 1 + \frac{1}{\e{\beta \omega_1'} - 1} \right) \left( 1 + \frac{1}{\e{\beta \omega_2'} - 1} \right) \left( 1 + \frac{1}{\e{\beta \omega_3'} - 1} \right) .
\end{split}
\end{equation}
\end{widetext}
While this expression appears rather involved, each factor has a clear physical interpretation. The combinatorial factor of $3!$ accounts for the symmetry of the three identical outgoing particles. The probability of doing a particular amount of work scales with volume of the system; the implications of this will be discussed shortly. The magnitude squared of the Fourier transform of the time-dependent coupling is the spectral density and can be thought of as a measure of how much the system is being driven in energy (frequency) space.

Finally we have a kinematic integral which is a function of the work performed. The integration measure is the Lorentz invariant momentum measure $\widetilde{d^{3} k}$ for each incoming and outgoing particle. The two sets of delta-functions impose conservation of energy and momentum including the contributions of the time-dependent background. The incoming particle is associated with the Bose-Einstein statistics factor, $1/(\e{\beta \omega} - 1)$. This is the density of states for a thermal system of bosons which should be expected because the system was initially prepared in thermal state. The outgoing particles, however, are associated with the unusual factor $1 + 1/(\e{\beta \omega} - 1)$. This is the appropriate density of states because the original occupancy number for a given energy level is just $1/(\e{\beta \omega} - 1)$ but due to the scattering process the occupancy of this level must go up by one.

These observations can be generalized to a set of rules for constructing any of the tree-level work distributions. One associates each incoming particle with the density of states $1/(\e{\beta \omega} - 1)$ and each outgoing particle with $1 + 1/(\e{\beta \omega} - 1)$. One then integrates over these kinematic factors in a Lorentz invariant manner and includes delta-functions for conserving energy and momentum. This is multiplied by the spectral density of the driving protocol and a factor of the volume. Appropriate symmetry factors for the incoming and outgoing particles are then included. In principle, one could arrive at these rules from a simple thermodynamic treatment of the density of states and subsequent use of classical field theory. We stress that this is not the approach that we used and that these expressions for the work distributions were derived by summing an infinite collection of Feynman diagrams in a fully quantum treatment.

As mentioned earlier, the work distribution function is proportional to the volume of the system. This leads to restrictions on the applicability of the work distributions in Appendix~\ref{app:Work-Distributions} to systems with large volume. As explained in Sec.~\ref{sec:Generic-Protocol}, unitary is not manifest at finite order in the Dyson series. This was sidestepped by noting that the Jarzynski equality still held for just the non-trivial component of the work distribution alone. However, it was assumed that the total work distribution function could be expressed as $\mc{P}(W) = a \, \delta(W) + \rho(W)$ where $a = 1 - \int dW \, \rho(W)$ is a positive constant and $\rho(W)$ is the non-trivial part of the work distribution. Since $\rho(W)$ is proportional to the volume, $a$ will become negative for large systems. At this point, our leading order approximation is no longer valid. Therefore, the range of validity of the present treatment is $\int dW \, \rho(W) < 1$. It may be possible, however, to extend the range of validity by working to higher orders in perturbation theory.

We now turn our attention to the component of the work distribution function which arises from loop diagrams, 
\begin{widetext}
\begin{equation}
\begin{aligned}
\rho^{\textrm{loop}}_{n \rightarrow n + 2} \left( W \right) = & \frac{V}{2} \, \left| \int dt \, \lambda(t) \e{i W t} \right|^2 \left( 1 + \frac{1}{\e{\beta W / 2} - 1} \right)^2 \frac{1}{W} \left( \int \widetilde{d^{3}k} \, \delta \left( W - 2 \omega \right) \right) \\
& \times \left( \int \widetilde{d^{3}k} \, \frac{1}{\e{\beta \omega} - 1} + \frac{1}{2} \left( \frac{m}{4 \pi} \right)^2 \left[ 1 + \log \left( \frac{\mu^2}{m^2} \right) \right] \right)^2 . \label{eq:loop-n-to-nplus2-mainbody}
\end{aligned}
\end{equation}
\end{widetext}
In this expression, $\mu$ is the renormalization scale in the $\overline{MS}$ scheme, see \textit{e.g.} Refs. \cite{Srednicki:1019751, Peskin:257493, Zee:706825}. Once again, we see that the work distribution is proportional to the volume and spectral density of the driving. In this process, there are two outgoing particles, each carrying half of the work put into the system which is reflected in $\left( 1 + 1/(\e{\beta W / 2} - 1) \right)^2$. The next two terms in \eqref{eq:loop-n-to-nplus2-mainbody} are a measure of the phase space available to the outgoing particles. It should be noted that because the particles have non-zero mass, $W \geq 2m$, one does not need to worry about the singular behavior of $1/W$. 

The final term in \eqref{eq:loop-n-to-nplus2-mainbody} results from the interference of two Feynman diagrams. The first diagram, shown in Fig.~\ref{fig:loop-n-to-nplus2}, is the one-loop process by which two particles can be created. This one-loop diagram, however, interferes with the tree-level diagram given in Fig.~\ref{fig:loop-tree-n-to-nplus2}. This tree-level process involves the production of two particles where the initial particle is merely a spectator and experiences no change in energy. The renormalization parameter $\mu$ then controls the relative size of the contribution from each diagram.

Loop diagrams do not exist in classical field theory and are the hallmark of second quantization. In a classical field theory Eq.~\eqref{eq:loop-n-to-nplus2-mainbody} would vanish and thus it may be thought of as the change in the work distribution function due to second quantization. Note, however, that $\mc{P}(W)$ may be dominated by these contributions, as we will see in  Sec.~\ref{sec:Numerical-Procotols}.

As the work distribution \eqref{eq:loop-n-to-nplus2-mainbody} explicitly depends on the renormalization scale $\mu$, this raises the question whether the work distribution is an observable quantity in quantum field theory. To be a physical, observable quantity, the work distribution should be independent of the renormalization scale and remain invariant under renormalization group flow. Beyond the explicit dependence on $\mu$, the coupling constant and mass have implicit dependence on $\mu$ due to renormalization. Using the $\beta$-function for this theory, it can be shown that the running of the work distribution enters at $\mathcal{O}\left( \lambda^3 \right)$. This is a higher order effect and may be modified by terms beyond leading order. To leading order the work distribution does not depend on the renormalization scale and therefore we conclude that the work distribution is, indeed, a physical observable.

\paragraph*{Fluctuation Theorems}
We now investigate how fluctuation theorems manifest in a quantum field theory. Throughout this section, we will always refer to the work distribution functions which are normalized such that the total work distribution integrates to unity. To make this normalization clear, we will utilize $P$ instead of $\rho$.

We first consider the Crooks fluctuation theorem \cite{Crooks1999,RevModPhys.83.771}. Assuming no change in free energy, the Crooks fluctuation theorem states that the probability distribution for a forward process, $P_{A \rightarrow B}(W)$, is related to the distribution for the reversed process, $P_{B \rightarrow A}(-W)$, through
\be
\frac{P_{B \rightarrow A}(-W) }{P_{A \rightarrow B}(W)} = \e{- \beta W}.
\ee

Accordingly, for the time-dependent field theory we have
\begin{equation}
\begin{split}
\frac{P_{n \rightarrow n - 4}(-W)}{P_{n \rightarrow n + 4}(W)} &= \frac{P_{n \rightarrow n - 2}(-W)}{P_{n \rightarrow n + 2}(W)}\\
= \frac{P_{n \rightarrow n}(-W)}{P_{n \rightarrow n}(W)} &= \e{-\beta W}.
\end{split}
\label{eq:Crooks-QFT}
\end{equation}
Using the explicit form of the work distribution functions from Appendix~\ref{app:Work-Distributions}, we verify \eqref{eq:Crooks-QFT} analytically. Again, this holds independently of the renormalization scale, $\mu$. The key property of the work distribution functions which allows for a proof of Eq.~\eqref{eq:Crooks-QFT} is that each incoming state is associated with the factor $1/(\e{\beta \omega} - 1)$ while each outgoing state is associated with $1 + 1/(\e{\beta \omega} - 1) = \e{\beta \omega}/(\e{\beta \omega} - 1)$. The latter is nothing else but an expression of local detailed balance.

More surprisingly, it can be shown that, independent of renormalization scale,
\be
\frac{P_{n \rightarrow n - 2}^{\textrm{tree}}(-W)}{P_{n \rightarrow n + 2}^{\textrm{tree}}(W)} = \frac{P_{n \rightarrow n - 2}^{\textrm{loop}}(-W)}{P_{n \rightarrow n + 2}^{\textrm{loop}}(W)} = \e{-\beta W}.
\ee
Thus, the Crooks fluctuation theorem holds both with and without the contribution from loop diagrams. Without loop diagrams, one simply has a classical field theory and the the validity of the fluctuation theorem is well established for classical systems. Moving to a quantum field theory, loop diagrams must be included but the Crooks fluctuation theorem still holds! This requires that order by order loop corrections must enter in a pairwise manner such that the fluctuation theorem holds at every order.

As the Crooks fluctuation theorem has been verified for our time-dependent field theory, the Jarzynski equality immediately follows as a consequence. This can quickly be shown through
\be
\int dW \, \mc{P}(W) \, \e{- \beta W} = \int dW \, \mc{P}(-W) = 1 .
\ee
As was true for the Crooks fluctuation theorem, the Jarzynski equality holds independent of the renormalization scale and will hold with or without the loop contributions.

In conclusion, we have analytically verified that the Crooks fluctuation theorem and Jarzynski equality hold independent of renormalization scale for a time-dependent quantum field theory at leading order. However, the quantum Jarzynski equality made no assumptions of perturbativity and only required the mild assumption of unital dynamics in Eqs.~\eqref{eq:Jarzynski-work-distribution} and \eqref{eq:quantum-Jarzynski}. Thus, while not verified analytically, these fluctuation theorems should hold to any order perturbatively and may even hold non-perturbatively.

\section{Example: numerical case study}
\label{sec:Numerical-Procotols}
We conclude the analysis with a numerical case study. Throughout this section, we will work in units such that $m=1$.
\begin{figure}[t]
	\subfloat[Normalized work distribution function. \label{fig:num-work-dist-total}]{
	\def\svgwidth{.42\textwidth}
	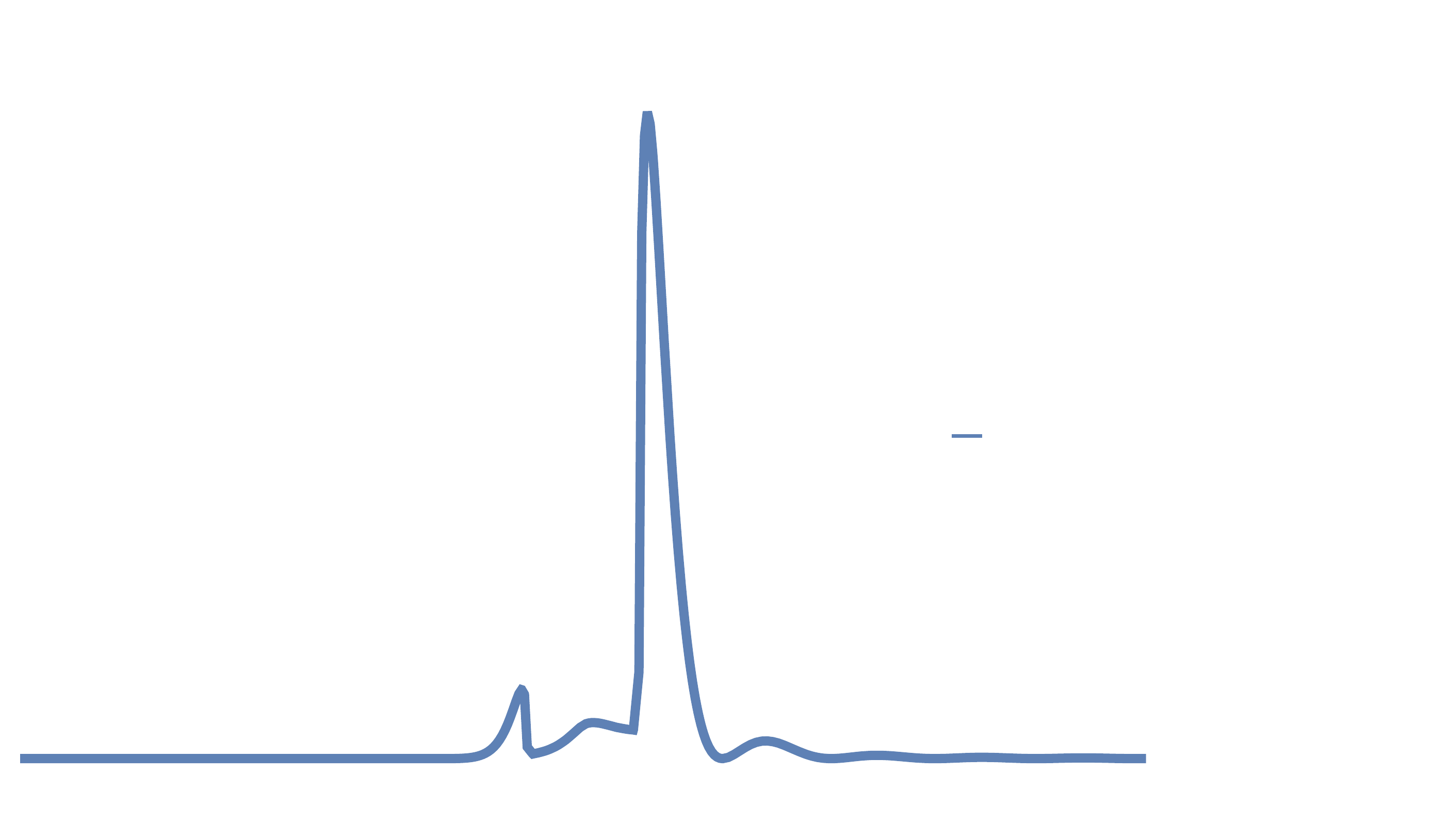
	} \newline
	\subfloat[Work distribution functions for subprocesses. \label{fig:num-work-dist-subs}]{
	\def\svgwidth{.42\textwidth}
	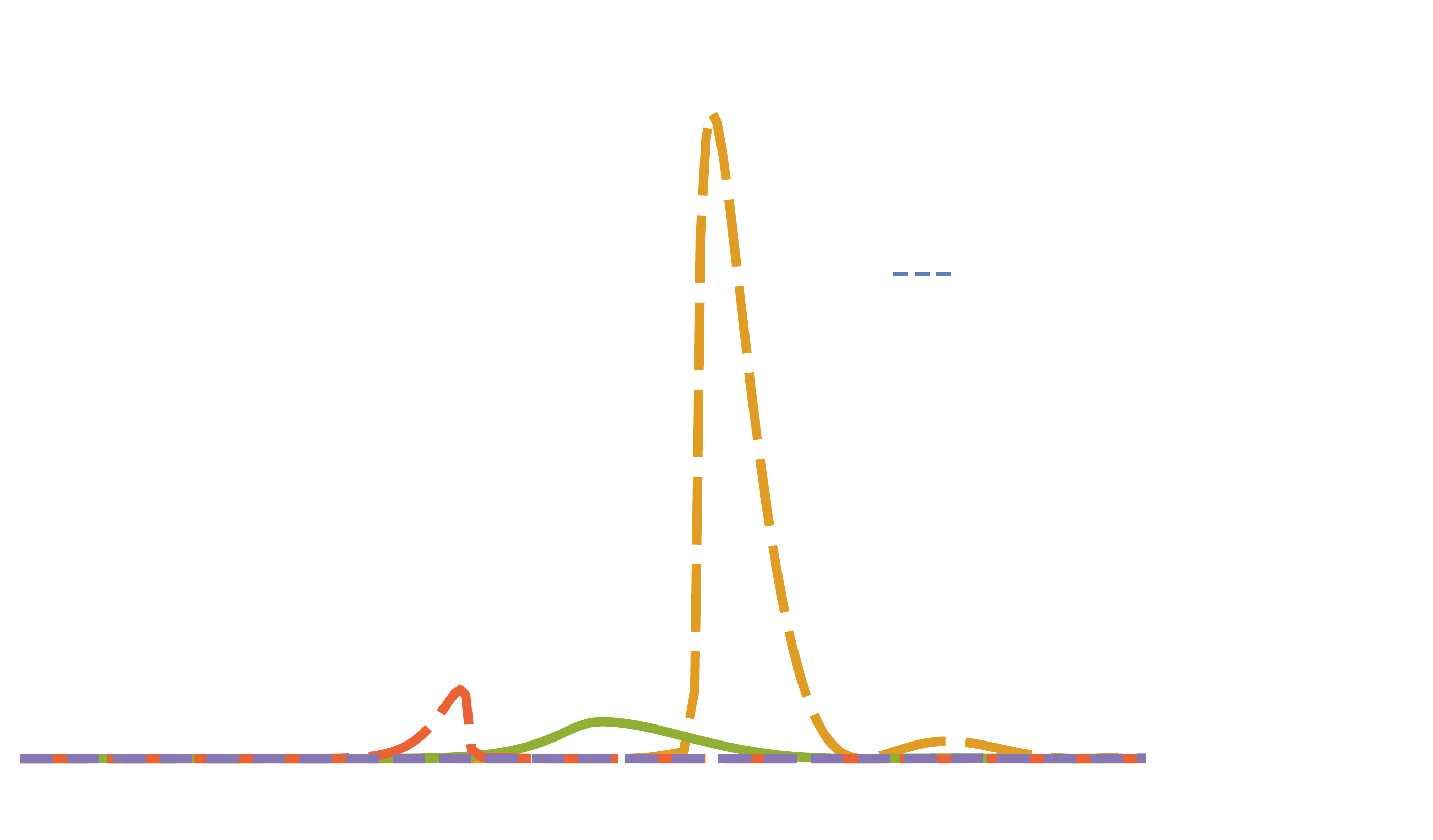
	} \newline
	\subfloat[Zoom-in of Fig.~\ref{fig:num-work-dist-subs} focusing on the tail of the work distributions. \label{fig:num-work-dist-zoom}]{
	\def\svgwidth{.42\textwidth}
	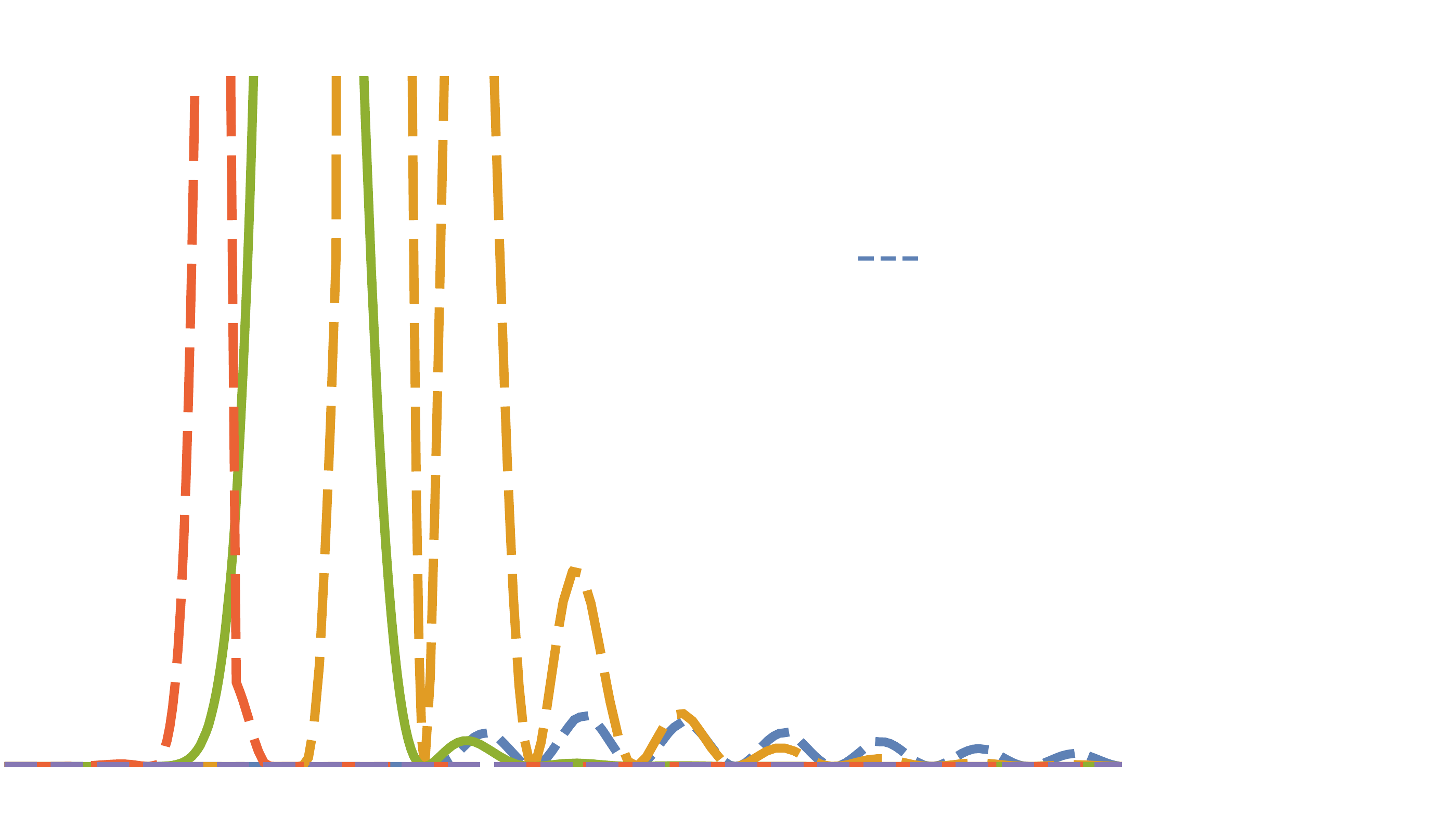
	}
	\caption{Work distribution function and its decomposition into subprocesses for $\beta = m = 1$ with driving protocol specified in \eqref{eq:driving-protocol}. \label{fig:num-work-dists}}
\end{figure}
To accentuate the contributions of particle creation and annihilation we will work with a relativistic bath and driving protocol. We will assume that the bath has inverse temperature comparable to the particle mass, $\beta = 1$. As driving protocol, we will consider the infinitely smooth but non-analytic ``bump'' function,
\be
	\lambda(t) = \left\{ \begin{array}{lr} \lambda_{0} \, \e{\frac{-t^2}{1-t^2}}, & \text{for } \left| t \right| \leq 1 \\
        							0, & \text{otherwise } \end{array} \right. .
\label{eq:driving-protocol}
\ee
This function is chosen to avoid any potential issues with continuity of derivatives at the start and end of the protocol. The overall scale of the driving protocol, $\lambda_{0}$, will ultimately drop out when the work distribution function is normalized. We only require that $\lambda_{0}$ is sufficiently small that the theory is perturbative and our expressions for the work distribution are valid.

We numerically evaluate the work distribution functions of Appendix~\ref{app:Work-Distributions} and subsequently normalize the combined $\mc{P}(W)$. This yields the total work distribution of Fig.~\ref{fig:num-work-dist-total} and the work distributions for the various subprocesses shown in Fig.~\ref{fig:num-work-dist-subs}. Both Fig.~\ref{fig:num-work-dist-total} and Fig.~\ref{fig:num-work-dist-subs} show the characteristic ``exponential asymmetry'' which is indicative of the Jarzynski equality. It is found that both the quantum Jarzynski equality and Crooks fluctuation theorem hold to within numerical precision. The dominant contribution to the work distribution function is from $P_{n \rightarrow n + 2}(W)$ with over $80\%$ of trials resulting in particle pair production. Surprisingly, within $P_{n \rightarrow n + 2}(W)$, over $95\%$ of the distribution is from the loop diagram contributions, $P^{\mathrm{loop}}_{n \rightarrow n + 2}(W)$. Thus, for this protocol and bath, the majority of the work distribution comes from loop diagrams which pair produce particles, an effect which only exists in a quantum field theory. Not including these loop diagrams would produce a markedly different $\mc{P}(W)$.

Figure~\ref{fig:num-work-dist-zoom} gives a zoomed in view of the tail of the work distribution functions. It can immediately be seen that all of the work distribution functions experience the same type of oscillatory behavior. This is a result of the spectral density of the time-dependent coupling vanishing at these energies (frequencies) and is not kinematic in origin. Figure~\ref{fig:num-work-dist-zoom} also shows that $P_{n \rightarrow n + 4}(W)$  is dominant over $P_{n \rightarrow n + 2}(W)$ but only at large values of work ($W \gtrsim 20$ mass units).

\section{Concluding remarks}
\label{sec:Conclusions}
Two decades ago, the Jarzynski equality was formulated for classical systems. It required another decade to formalize the concept of work in the quantum regime. However, quantum mechanics is not our most complete description of nature. This is quantum field theory, and another decade later quantum fluctuation theorems have now been extended to their ultimate limit.

Considering a time-dependent variant of $\lambda \phi^4$, we found closed form expressions for the work distribution functions. While this is only one particular quantum field theory, these distribution functions demonstrate a variety of features which should be anticipated in other field theories. It was shown that the work distribution functions do not run with the renormalization scale through the perturbative order considered, implying that these distributions are physical observables of the field theory. It was also found that both the Crooks fluctuation theorem and Jarzynski equality hold independent of the renormalization scale, both with and without loop corrections. Remarkably, the contribution of loop diagrams to the work distribution function can be substantial and thus essential in the proper description of work fluctuations of quantum field theories in the relativistic regime.

Until now, particle pair-production and loop effects had not been incorporated in any study of quantum fluctuation theorems. These effects become dominant in the relativistic regime. Our results were presented in units of the particle mass which can obscure physical intuition for the system and protocol being considered. To get better insight, consider a hypothetical condensed matter system that is described by the time-dependent $\lambda \phi^4$ theory with an effective mass $m \sim 1 \, \mathrm{eV}$. Our work distributions describe the behavior of such a system with an effective temperature of $T \sim 10^6 \, \mathrm{K}$ with a driving time of $\Delta t = 10^{-15} \, \mathrm{s}$. For a particle with mass comparable to the electron, $m \sim 1 \, \mathrm{MeV}$, this corresponds to temperatures of $T \sim 10^{12} \, \mathrm{K}$ and driving times of $\Delta t = 10^{-21} \, \mathrm{s}$; conditions relevant for the study of quark-gluon plasma. This is well beyond the original regime in which the fluctuation theorems were conceived and outside previous treatments in the literature.

It should be stressed that these results were directly calculated from in-out scattering amplitudes which are the natural building block for the quantum Jarzynski equality. In particular, we did not need to utilize the Schwinger-Keldysh in-in formalism at any point in the calculation. This was only possible because we were able to find a mapping between finite-time transition amplitudes and infinite-time amplitudes, and were able to show that disconnected vacuum diagrams did not alter the scattering amplitude. Due to the form of the work distribution function a new diagrammatic technique had to be developed so that the infinite sum over particle number and sum over permutations of disconnected Feynman diagrams can be performed analytically. This technique relies on the topological properties of ``glued Feynman diagrams'' to classify permutations and enables the rephrasing of the sum over particle number in terms of a sum over graph theoretic properties of the glued diagrams.

While this work focused primarily on time-dependent $\lambda \phi^4$, these techniques should be applicable to any quantum field theory. Even with the restriction that the system begins and ends as a free field theory, this vastly expands the realm of applicability for quantum fluctuation theorems. One example of an interesting system which fits within this paradigm is a cyclic engine acting on a quantum field working medium. How particle pair-production effects the work distribution of such an engine is an unstudied problem which should now be addressable given the techniques outlined in this work.

This opens new frontiers for the use of fluctuation theorems; from the relativistic charge carriers of graphene to the quark gluon plasma produced in heavy ion collidors to the evolution of the early universe, fluctuation theorems can provide insight into the short timescale behavior of nonequilibrium systems. While we may now begin applying quantum fluctuation theorems to the most extreme conditions found in nature, there is still more progress to be made. The most immediate challenge is to generalize the approach presented here to make it applicable to a wider variety of quantum field theories. This would enable the study of gauge fields and more interesting protocols. Twenty years after the advent of the Jarzynski equality and ten years after its quantum equivalent, fluctuation theorems can finally be applied across the full range of energy and length scales understood in modern physics but more work is still left to be done.

\acknowledgements{The authors would like to thank the stimulating environment provided by the Telluride Science Research Center, where this project was conceived. AB would like to thank Mark Wise and Sean Carroll for helpful discussions on time-dependent field theory. AB acknowledges support from the U.S. Department of Energy, Office of Science, Office of High Energy Physics, under Award Number DE-SC0011632. SD acknowledges support from the U.S. National Science Foundation under Grant No. CHE-1648973.}{

\begin{appendix}
\begin{widetext}

\section{Finite-time Amplitude to Infinite-time Amplitude Mapping}
\label{app:Finite-to-Infinite}

In calculating the work distribution function \eqref{eq:Work-Dist-Scattering} for a quantum field theory, one must address the finite-time transition amplitude $\left \langle k_1 ', \ldots, k_{n_2}' ; t_2 \right| U(t_2, t_1) \left| k_1, \ldots, k_{n_1}; t_1 \right\rangle$. This finite-time amplitude must be rewritten in terms of an infinite-time scattering process so that the full machinery of quantum field theory can be used.

In general, the initial and final states of the system will be multiparticle states, however we will specialize to the case of single particle states for notational simplicity. The generalization to multiparticle states is straightforward.

To distinguish operators in different quantum mechanical pictures, all states and operators in the Schr\"odinger picture will be denoted by a subscript $S$ and those in the Interaction picture will have a subscript $I$. We start by defining the initial and final states in terms of the creation and annihilation operators as $\left| k; t_1 \right\rangle_{S} = a_{S}^{\dagger}(k) \left| \Phi_{\mathrm{in}}; t_{1} \right\rangle_{S}$ and $\left| k' ; t_2 \right\rangle_{S} = a_{S}^{\dagger}(k') \left| \Phi_{\mathrm{out}}; t_{2} \right\rangle_{S}$. In these expressions, $\left| \Phi_{\mathrm{in}}; t_{1} \right\rangle_{S}$ and $\left| \Phi_{\mathrm{out}}; t_{2} \right\rangle_{S}$ are the in-coming and out-going vacuum states. These states are defined such that $H_{0} \left| \Phi_{\mathrm{in}}; t_{1} \right\rangle_{S} = H_{0} \left| \Phi_{\mathrm{out}}; t_{2} \right\rangle_{S} = 0$. Using these definitions,
\begin{equation}
{}_{S} \left \langle k' ; t_2 \right| U(t_2, t_1) \left| k ; t_1 \right\rangle_{S} =
{}_{S}\left\langle \Phi_{\mathrm{out}}; t_{2} \right| a_{S}(k') U(t_{2}, t_{1}) a^{\dagger}_{S}(k) \left| \Phi_{\mathrm{in}}; t_{1} \right\rangle_{S}.
\end{equation}

We now define a time $\tau \gg \max\left( \left| t_1 \right| ,\left| t_2 \right| \right)$ with the intention of eventually taking the limit $\tau \rightarrow \infty$. As the system is assumed to be free at times $t_1$ and $t_2$, one may trivially extend the finite-time experiment by assuming that the system remains free outside of the interval $t\in \left( t_1, t_2 \right)$. This cannot change the work distribution function as no work is performed keeping the Hamiltonian fixed. Then,
\be
	{}_{S} \left\langle k' ; t_{2} \right| U(t_{2}, t_{1}) \left| k ; t_{1} \right\rangle_{S} = {}_{S}\left\langle \Phi_{\mathrm{out}}; \tau \right| U(\tau,t_{2}) a_{S}(k') U(t_{2}, t_{1}) a^{\dagger}_{S}(k) U(t_{1},-\tau) \left| \Phi_{\mathrm{in}}; -\tau \right\rangle_{S} . 
\ee
Note, as $H(t) = H_0$ for $t \notin \left( t_1, t_2 \right)$, it is still true that $H_{0} \left| \Phi_{\mathrm{in}}; -\tau \right\rangle_{S} = H_{0} \left| \Phi_{\mathrm{out}}; \tau \right\rangle_{S} = 0$.

We now define some reference time $t_{0} \notin \left(t_{1}, t_{2} \right)$ when the Schr\"odinger and Interaction pictures coincide. Passing to the Interaction picture,
\be
{}_{S} \left\langle k' ; t_{2} \right| U(t_{2}, t_{1}) \left| k ; t_{1} \right\rangle_{S} =
{}_{S}\left\langle \Phi_{\mathrm{out}}; \tau \right| U_{0}( \tau, t_{0}) U_{I}(\tau, t_{2}) a_{I}(k'; t_2) U_{I}(t_{2}, t_{1}) a^{\dagger}_{I}(k; t_1) U_{I}(t_{1},-\tau) U_{0}( t_{0}, -\tau) \left| \Phi_{\mathrm{in}}; -\tau \right\rangle_{S}.
\ee
In this expression, $U_{0}$ is the time evolution operator under the free Hamiltonian while $U_{I}$ is the evolution operator in the Interaction picture. As $\left| \Phi_{\mathrm{out}}; \tau \right \rangle_{S}$ and $\left| \Phi_{\mathrm{in}}; -\tau \right\rangle_{S}$ are vacuum states of the free theory, ${}_{S}\left\langle \Phi_{\mathrm{out}}; \tau \right| U_{0}( \tau, t_{0}) = {}_{S}\left\langle \Phi_{\mathrm{out}}; \tau \right|$ and $U_{0}( t_{0}, -\tau) \left| \Phi_{\mathrm{in}}; -\tau \right\rangle_{S} = \left| \Phi_{\mathrm{in}}; -\tau \right\rangle_{S}$. Furthermore, as both states are annihilated by the free Hamiltonian and the ground state is unique, they may differ by at most a phase from the ground state $\left| \Omega; t_{0} \right \rangle_{I}$. As the scattering amplitude will be squared in the final calculation, these phase factors are irrelevant. Rewriting the scattering amplitude as a time-ordered product,
\be
	\left| {}_{S} \left\langle k' ; t_{2} \right| U(t_{2}, t_{1}) \left| k ; t_{1} \right\rangle_{S} \right| = \left| {}_{I}\left\langle \Omega \right| T \left[ U_{I}(\tau,-\tau) a_{I}(k'; t_2) a^{\dagger}_{I}(k; t_1) \right] \left| \Omega\right\rangle_{I} \right| .
\label{eq:time-ordered-creation}
\ee

Using the mode-expansion of the free scalar field and the definition of time-evolution for operators in the Interaction picture, it is straightforward to show,
\begin{align}
	\e{i \omega t} a_{I} \left( k; t \right) &= i \int d^{3}x \; \e{-i k x} \doubleoverarrow{\partial_{0}} \phi_{I}\left( x \right) , \label{eq:annihilation-Schro-Int}\\
	\e{-i \omega t} a^{\dagger}_{I} \left( k; t \right) &= -i \int d^{3}x \; \e{i k x} \doubleoverarrow{\partial_{0}} \phi_{I}\left( x \right) .\label{eq:creation-Schro-Int}
\end{align}
In these relations, the operator $\doubleoverarrow{\partial_{\mu}}$ is defined such that 
$f \doubleoverarrow{\partial_{\mu}} g = f \left( \partial_{\mu} g \right) - \left( \partial_{\mu} f \right) g$, see Ref. \cite{Srednicki:1019751}.

Making use of \eqref{eq:annihilation-Schro-Int} and \eqref{eq:creation-Schro-Int}, it is possible to rewrite Eq.~\eqref{eq:time-ordered-creation} purely in terms of field operators in the Interaction picture. As noted before, we are only interested in the magnitude of Eq.~\eqref{eq:time-ordered-creation} as any overall phase disappears in Eq.~\eqref{eq:Work-Dist-Scattering}. Thus, up to an overall irrelevant phase, we find
\be
	\left| {}_{S} \left\langle k' ; t_{2} \right| U(t_{2}, t_{1}) \left| k ; t_{1} \right\rangle_{S} \right| = \left| \int d^{3}x' d^{3}x \; \e{-i k' x'} \e{i k x} \doubleoverarrow{\partial_{0_{x'}}} \doubleoverarrow{\partial_{0_{x}}} \, {}_{I}\left\langle \Omega \right| T \left[ U_{I}(\tau,-\tau) \phi_{I}(x') \phi_{I}(x) \right] \left| \Omega\right\rangle_{I} \right| .
\ee
In this expression, it is understood that the time components of the four-vectors $x$ and $x'$ are to be evaluated at $t_{1}$ and $t_{2}$ respectively.

We may now formally take the limit as $\tau \rightarrow \infty$. Generalizing to the case of multiparticle initial and final states,
\begin{equation}
\begin{split}
&\left| \left \langle k_1 ', \ldots, k_{n_2}' \right| U(t_2, t_1) \left| k_1, \ldots, k_{n_1} \right\rangle \right| = \\
&\quad \left| \int \right. d^{3}x_{1}' d^{3}x_{1} \ldots \; \e{-i k_{1}' x_{1}'} \e{i k_{1} x_{1}}\ldots\doubleoverarrow{\partial_{0_{x_{1}'}}} \doubleoverarrow{\partial_{0_{x_{1}}}} \ldots \left. \cdot {}_{I}\left\langle \Omega \right| T \left[ U_{I}(\infty,-\infty) \phi_{I}(x_1 ') \ldots \phi_{I}(x_1) \ldots \right] \left| \Omega \right\rangle_{I} \vphantom{\int} \right|.
\end{split}
\label{eq:Finite-Time-To-QFT}
\end{equation}

\section{Diagrammatic Technique}
\label{app:Techniques}
As mentioned in Sec.~\ref{sec:Generic-Protocol}, when calculating the work distribution function from the Dyson series one runs into technical difficulties. The interaction Hamiltonian \eqref{eq:Interaction-Hamiltonian} is composed of terms which involve at most four field sources while the general work distribution function \eqref{eq:Work-Dist-Scattering} involves any number of incoming or outgoing particles. As a result the Feynman diagrams which describe these processes will be composed of several disconnected subdiagrams. One must sum over all possible permutations of these subdiagrams before squaring the resulting amplitude; unlike the usual procedure in quantum field theory where one is only interested in fully connected diagrams and their permutations.

To motivate our procedure for calculating this sum, it will be necessary to introduce notation for describing the permutations of Feynman diagrams. From Eq.~\eqref{eq:Work-Dist-Scattering} it can be seen that the scattering amplitude will depend on the momenta of the incoming and outgoing particles and all of these momenta are integrated over in a Lorentz invariant manner. Let $K$ denote the collection of all momenta and let $dK$ denote the Lorentz invariant measure. We will let $\mathcal{D}$ correspond to a Feynman diagram of interest, such as the one shown in Fig.~\ref{fig:disconnected-unperm}. Let $\mathcal{S}$ be the set of all permutations of the Feynman diagram which do not interchange incoming for outgoing particles and let $\sigma \in \mathcal{S}$ be a particular permutation. Now define $f \left( W, K \right)$ to be all the terms that appear in Eq.~\eqref{eq:Work-Dist-Scattering} that are not the scattering amplitude, \textit{i.e.} the energy conserving delta-function, Boltzmann factors, and normalization constant. Then, Eq.~\eqref{eq:Work-Dist-Scattering} can schematically be rewritten as
\begin{align}
P(W) &= \int d K \, f\left( W, K \right) \cdot \left| \sum_{\sigma \in \mathcal{S}} \left( \sigma \circ \mathcal{D} \right) \left( K \right) \right|^2 \label{eq:perms-square-of-sum} \\
&= \sum_{\sigma_1, \sigma_2 \in \mathcal{S}} \int d K \, f\left( W, K \right) \cdot \left( \sigma_1 \circ \mathcal{D} \right)^{\dagger} \left( K \right) \cdot \left( \sigma_{2} \circ \mathcal{D} \right) \left( K \right) \label{eq:perms-sum-of-products}\\
&= \sum_{\sigma_1, \sigma_2 \in \mathcal{S}} \int d K \, f\left( W, K \right) \cdot \mathcal{D}^{\dagger} \left( K \right) \cdot \left( \sigma_{2} \circ \sigma_{1}^{-1} \circ  \mathcal{D} \right) \left( K \right) \label{eq:perms-relabelling} \\
&= \left| \mathcal{S} \right| \sum_{\sigma \in \mathcal{S}} \int d K \, f\left( W, K \right) \cdot \mathcal{D}^{\dagger} \left( K \right) \cdot \left( \sigma \circ \mathcal{D} \right) \left( K \right) , \label{eq:perms-final}
\end{align}
where $\left| \mathcal{S} \right|$ is the total number of permutations of the diagram $\mathcal{D}$. In moving from Eq.~\eqref{eq:perms-square-of-sum} to Eq.~\eqref{eq:perms-sum-of-products} we have rewritten the square of the sum as the sum over crossterms. In Eq.~\eqref{eq:perms-relabelling} we have chosen to relabel the momenta $K$ such that the first diagram $\mathcal{D}^{\dagger}$ appears unpermuted. Lastly, in Eq.~\eqref{eq:perms-final} we have identified the composition of permutations as a permutation and performed the sum over the redundant permutation. In these expressions, we have suppressed the sum over incoming and outgoing particle number and have ignored potential complications arising from a process mediated by more than one type of Feynman diagram. Using Eq.~\eqref{eq:perms-final}, the work distribution function can be calculated by integrating over the momenta of the product of an ``unpermuted'' Feynman diagram and its possible permutations.

Since we are studying a variant of $\lambda \phi^4$, at leading order the particle number may either stay the same, change by two, or change by four. Diagrams with different numbers of incoming or outgoing particles do not interfere and thus can be considered separately, as mentioned in Sec.~\ref{sec:Generic-Protocol}. For concreteness, we will now consider processes involving $n$ particles where the particle number is unchanged.

For processes where the particle number is unchanged, the only Feynman diagrams that contribute at leading order are permutations of Fig.~\ref{fig:disconnected-unperm}. This particular diagram is drawn for $n=6$ and for clarity the insertion of the background field $\chi_{\mathrm{cl}}$ is not shown. We will choose this diagram to represent the ``unpermuted'' Feynman diagram $\mathcal{D}$. In principle, any other valid diagram could be chosen as the ``unpermuted'' reference, but Fig.~\ref{fig:disconnected-unperm} is chosen for convenience. Three possible permutations of this diagram are shown in Fig.~\ref{fig:disconnected-perms}. Note that these permutations only interchange incoming particles amongst themselves or outgoing particles amongst themselves. It should also be noted that the exchange $1 \leftrightarrow 2$ is not considered a unique permutation because it leaves the overall diagram unchanged.

Before Eq.~\eqref{eq:perms-final} may be utilized to calculate the work distribution function, one needs to define a scheme for enumerating possible permutations of Fig.~\ref{fig:disconnected-unperm}. It will now be demonstrated that the three permutations shown in Figs.~\ref{fig:disconnected-perm1},~\ref{fig:disconnected-perm2}, and~\ref{fig:disconnected-perm3} define three classes of permutation which will uniquely catagorize any permutation of Fig.~\ref{fig:disconnected-unperm}. 
\begin{figure*}[t]
	\hspace*{\fill}
	\subfloat[The prototypical example for a Feynman diagram which contributes to $n \rightarrow n$ scattering at leading order. \label{fig:disconnected-unperm}]{
	\begin{fmffile}{fgraphs09}
		\begin{fmfgraph*}(50,50)
		\fmfset{thick}{1.25}
		\fmfpen{thick}
		\fmfforce{(.1w,.95h)}{i1} 
		\fmfforce{(.1w,.80h)}{i2} 
		\fmfforce{(.1w,.65h)}{i3} 
		\fmfforce{(.1w,.50h)}{i4} 
		\fmfforce{(.1w,.35h)}{i5} 
		\fmfforce{(.1w,.20h)}{i6} 
		\fmfforce{(.9w,.95h)}{o1} 
		\fmfforce{(.9w,.80h)}{o2} 
		\fmfforce{(.9w,.65h)}{o3} 
		\fmfforce{(.9w,.50h)}{o4} 
		\fmfforce{(.9w,.35h)}{o5} 
		\fmfforce{(.9w,.20h)}{o6}
		\fmf{plain, tension=1}{i1,v1}
		\fmf{plain, tension=1}{i2,v1}
		\fmf{plain, tension=1}{v1,o1}
		\fmf{plain, tension=1}{v1,o2}
		\fmf{dashes, tension=1}{i3,o3}
		\fmf{dashes, tension=1}{i4,o4}
		\fmf{dashes, tension=1}{i5,o5}
		\fmf{dashes, tension=1}{i6,o6}
		\fmfdot{v1}
		\fmflabel{\large$1$}{i1}
		\fmflabel{\large$2$}{i2}
		\fmflabel{\large$3$}{i3}
		\fmflabel{\large$4$}{i4}
		\fmflabel{\large$5$}{i5}
		\fmflabel{\large$6$}{i6}
		\fmflabel{\large$1'$}{o1}
		\fmflabel{\large$2'$}{o2}
		\fmflabel{\large$3'$}{o3}
		\fmflabel{\large$4'$}{o4}
		\fmflabel{\large$5'$}{o5}
		\fmflabel{\large$6'$}{o6}
		\end{fmfgraph*}
	\end{fmffile}
	}
	\hspace*{\fill}
	\subfloat[One type of permutation of the Feynman diagram in Fig.~\ref{fig:disconnected-unperm}. \label{fig:disconnected-perm1}]{
	\begin{fmffile}{fgraphs10}
		\begin{fmfgraph*}(50,50)
		\fmfset{thick}{1.25}
		\fmfpen{thick}
		\fmfforce{(.1w,.95h)}{i1} 
		\fmfforce{(.1w,.80h)}{i2} 
		\fmfforce{(.1w,.65h)}{i3} 
		\fmfforce{(.1w,.50h)}{i4} 
		\fmfforce{(.1w,.35h)}{i5} 
		\fmfforce{(.1w,.20h)}{i6} 
		\fmfforce{(.9w,.95h)}{o1} 
		\fmfforce{(.9w,.80h)}{o2} 
		\fmfforce{(.9w,.65h)}{o3} 
		\fmfforce{(.9w,.50h)}{o4} 
		\fmfforce{(.9w,.35h)}{o5} 
		\fmfforce{(.9w,.20h)}{o6}
		\fmf{plain, tension=1}{i3,v1}
		\fmf{plain, tension=1}{i6,v1}
		\fmf{plain, tension=1}{v1,o4}
		\fmf{plain, tension=1}{v1,o5}
		\fmf{dashes, tension=1}{i1,o1}
		\fmf{dashes, tension=1}{i2,o2}
		\fmf{dashes, tension=1}{i4,o3}
		\fmf{dashes, tension=1}{i5,o6}
		\fmfdot{v1}
		\fmflabel{\large$1$}{i1}
		\fmflabel{\large$2$}{i2}
		\fmflabel{\large$3$}{i3}
		\fmflabel{\large$4$}{i4}
		\fmflabel{\large$5$}{i5}
		\fmflabel{\large$6$}{i6}
		\fmflabel{\large$1'$}{o1}
		\fmflabel{\large$2'$}{o2}
		\fmflabel{\large$3'$}{o3}
		\fmflabel{\large$4'$}{o4}
		\fmflabel{\large$5'$}{o5}
		\fmflabel{\large$6'$}{o6}
		\end{fmfgraph*}
	\end{fmffile}
	}
	\hspace*{\fill}
	\newline
	\hspace*{\fill}
	\subfloat[A second possible permutation of the Feynman diagram in Fig.~\ref{fig:disconnected-unperm}. \label{fig:disconnected-perm2}]{
	\begin{fmffile}{fgraphs11}
		\begin{fmfgraph*}(50,50)
		\fmfset{thick}{1.25}
		\fmfpen{thick}
		\fmfforce{(.1w,.95h)}{i1} 
		\fmfforce{(.1w,.80h)}{i2} 
		\fmfforce{(.1w,.65h)}{i3} 
		\fmfforce{(.1w,.50h)}{i4} 
		\fmfforce{(.1w,.35h)}{i5} 
		\fmfforce{(.1w,.20h)}{i6} 
		\fmfforce{(.9w,.95h)}{o1} 
		\fmfforce{(.9w,.80h)}{o2} 
		\fmfforce{(.9w,.65h)}{o3} 
		\fmfforce{(.9w,.50h)}{o4} 
		\fmfforce{(.9w,.35h)}{o5} 
		\fmfforce{(.9w,.20h)}{o6}
		\fmf{plain, tension=1}{i3,v1}
		\fmf{plain, tension=1}{i6,v1}
		\fmf{plain, tension=1}{v1,o4}
		\fmf{plain, tension=1}{v1,o5}
		\fmf{dashes, tension=1}{i1,o1}
		\fmf{dashes, tension=1}{i2,o3}
		\fmf{dashes, tension=1}{i4,o2}
		\fmf{dashes, tension=1}{i5,o6}
		\fmfdot{v1}
		\fmflabel{\large$1$}{i1}
		\fmflabel{\large$2$}{i2}
		\fmflabel{\large$3$}{i3}
		\fmflabel{\large$4$}{i4}
		\fmflabel{\large$5$}{i5}
		\fmflabel{\large$6$}{i6}
		\fmflabel{\large$1'$}{o1}
		\fmflabel{\large$2'$}{o2}
		\fmflabel{\large$3'$}{o3}
		\fmflabel{\large$4'$}{o4}
		\fmflabel{\large$5'$}{o5}
		\fmflabel{\large$6'$}{o6}
		\end{fmfgraph*}
	\end{fmffile}
	} 
	\hspace*{\fill}
	\subfloat[A third permutation of the Feynman diagram in Fig.~\ref{fig:disconnected-unperm}. \label{fig:disconnected-perm3}]{
	\begin{fmffile}{fgraphs12}
		\begin{fmfgraph*}(50,50)
		\fmfset{thick}{1.25}
		\fmfpen{thick}
		\fmfforce{(.1w,.95h)}{i1} 
		\fmfforce{(.1w,.80h)}{i2} 
		\fmfforce{(.1w,.65h)}{i3} 
		\fmfforce{(.1w,.50h)}{i4} 
		\fmfforce{(.1w,.35h)}{i5} 
		\fmfforce{(.1w,.20h)}{i6} 
		\fmfforce{(.9w,.95h)}{o1} 
		\fmfforce{(.9w,.80h)}{o2} 
		\fmfforce{(.9w,.65h)}{o3} 
		\fmfforce{(.9w,.50h)}{o4} 
		\fmfforce{(.9w,.35h)}{o5} 
		\fmfforce{(.9w,.20h)}{o6}
		\fmf{plain, tension=1}{i3,v1}
		\fmf{plain, tension=1}{i6,v1}
		\fmf{plain, tension=1}{v1,o4}
		\fmf{plain, tension=1}{v1,o5}
		\fmf{dashes, tension=1}{i1,o3}
		\fmf{dashes, tension=1}{i2,o6}
		\fmf{dashes, tension=1}{i4,o1}
		\fmf{dashes, tension=1}{i5,o2}
		\fmfdot{v1}
		\fmflabel{\large$1$}{i1}
		\fmflabel{\large$2$}{i2}
		\fmflabel{\large$3$}{i3}
		\fmflabel{\large$4$}{i4}
		\fmflabel{\large$5$}{i5}
		\fmflabel{\large$6$}{i6}
		\fmflabel{\large$1'$}{o1}
		\fmflabel{\large$2'$}{o2}
		\fmflabel{\large$3'$}{o3}
		\fmflabel{\large$4'$}{o4}
		\fmflabel{\large$5'$}{o5}
		\fmflabel{\large$6'$}{o6}
		\end{fmfgraph*}
	\end{fmffile}
	}
	\hspace*{\fill}
	\newline
	\caption{An incomplete collection of possible Feynman diagrams which contribute to the non-trivial part of the work distribution for $n \rightarrow n$ scattering. For visual clarity, propogators which are not part of the four-point function are represented with dashed lines and insertions of the background field $\chi_{\mathrm{cl}}$ are omitted. \label{fig:disconnected-perms}}
\end{figure*}
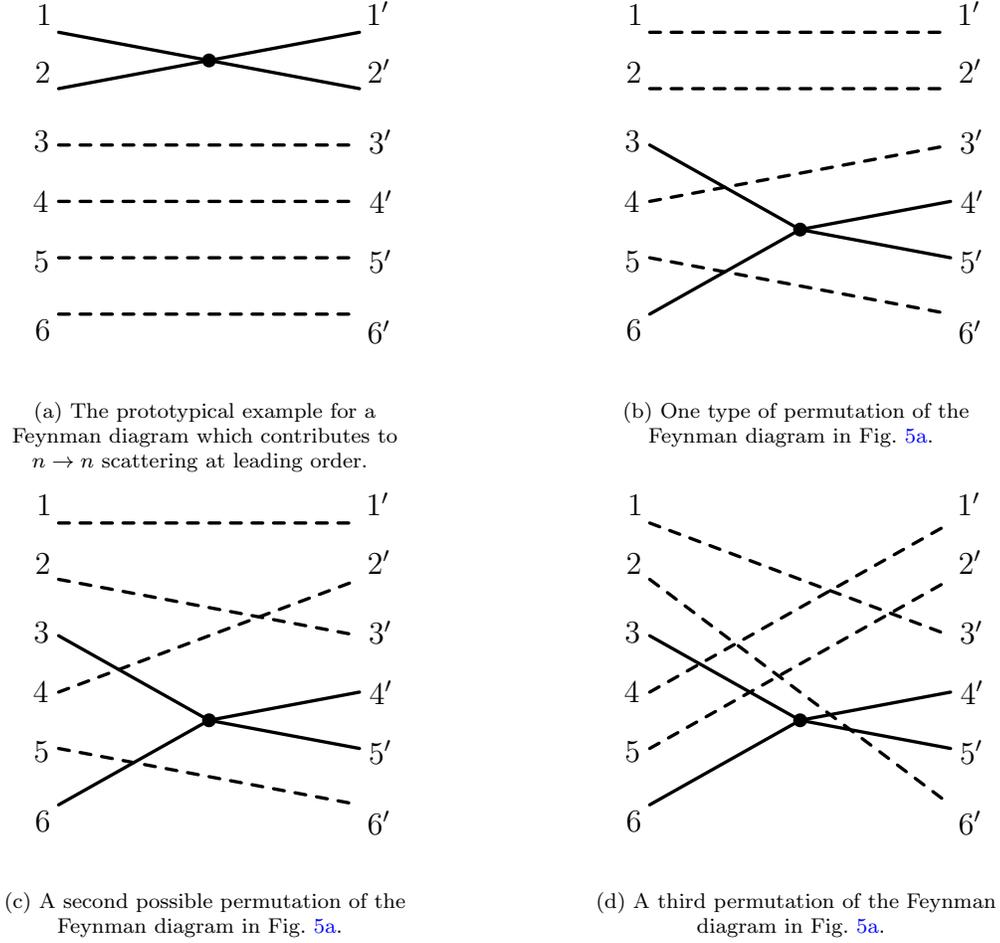

Consider Eq.~\eqref{eq:perms-final}. One is interested in the product of two Feynman diagrams: $\mathcal{D}^{\dagger} \left( K \right)$, the conjugate of the unpermuted diagram, and $\left( \sigma \circ \mathcal{D} \right) \left( K \right)$, a permuted Feynman diagram. In each Feynman diagram, the incoming and outgoing momenta are the same. It is only how these momenta are connected to one another through delta-functions and four-point functions which differs. As the momenta are identical, it will be helpful to define a ``glued'' Feynman diagram which is built from the two Feynman diagrams by treating the field sources for each diagram as identical. For example, in Figs.~\ref{fig:disconnected-unperm} and~\ref{fig:disconnected-perm1}, the momentum associated with particle $1$ in each diagram is the same. As such, these diagrams can be connected by ``gluing'' the diagrams together at this point. Repeating this for each incoming and outgoing field source yields Fig.~\ref{fig:glued-case1}. For each permuted diagram in Figure~\ref{fig:disconnected-perms}, the corresponding ``glued diagram'' is shown in Figure~\ref{fig:glued-diagrams}.
\begin{figure*}[t]
	\hspace*{\fill}
	\subfloat[Glued diagram resulting from the combination of Fig.~\ref{fig:disconnected-unperm} and Fig.~\ref{fig:disconnected-perm1}. \label{fig:glued-case1}]{
	\begin{fmffile}{fgraphs13}
		\begin{fmfgraph*}(50,50)
		\fmfset{thick}{1.25}
		\fmfpen{thick}
		\fmfforce{(.1w,.90h)}{i1} 
		\fmfforce{(.3w,.90h)}{o1} 
		\fmfforce{(.3w,.15h)}{i2} 
		\fmfforce{(.1w,.15h)}{o2} 
		\fmfforce{(.9w,.90h)}{i3} 
		\fmfforce{(.5w,.90h)}{o4} 
		\fmfforce{(.5w,.15h)}{i6} 
		\fmfforce{(.9w,.15h)}{o5} 
		\fmf{plain, tension=1}{i1,v1}
		\fmf{plain, tension=1}{i2,v1}
		\fmf{plain, tension=1}{v1,o1}
		\fmf{plain, tension=1}{v1,o2}
		\fmf{plain, tension=1}{i3,v2}
		\fmf{plain, tension=1}{i6,v2}
		\fmf{plain, tension=1}{v2,o4}
		\fmf{plain, tension=1}{v2,o5}
		\fmf{dashes, tension=1}{i1,o1}
		\fmf{dashes, tension=1}{i2,o2}
		\fmf{dashes, tension=1}{i3,o3}
		\fmf{dashes, tension=1}{o3,i4}
		\fmf{dashes, tension=1}{i4,o4}
		\fmf{dashes, tension=1}{i6,o6}
		\fmf{dashes, tension=1}{o6,i5}
		\fmf{dashes, tension=1}{i5,o5}
		\fmfdot{v1}
		\fmfdot{v2}
		\fmfv{decor.shape=square,decor.filled=full,decor.size=2.5thick}{i1}
		\fmfv{decor.shape=square,decor.filled=full,decor.size=2.5thick}{i2}
		\fmfv{decor.shape=square,decor.filled=full,decor.size=2.5thick}{i3}
		\fmfv{decor.shape=square,decor.filled=full,decor.size=2.5thick}{i4}
		\fmfv{decor.shape=square,decor.filled=full,decor.size=2.5thick}{i5}
		\fmfv{decor.shape=square,decor.filled=full,decor.size=2.5thick}{i6}
		\fmfv{decor.shape=square,decor.filled=full,decor.size=2.5thick}{o1}
		\fmfv{decor.shape=square,decor.filled=full,decor.size=2.5thick}{o2}
		\fmfv{decor.shape=square,decor.filled=full,decor.size=2.5thick}{o3}
		\fmfv{decor.shape=square,decor.filled=full,decor.size=2.5thick}{o4}
		\fmfv{decor.shape=square,decor.filled=full,decor.size=2.5thick}{o5}
		\fmfv{decor.shape=square,decor.filled=full,decor.size=2.5thick}{o6}
		\fmflabel{\large$1$}{i1}
		\fmflabel{\large$2$}{i2}
		\fmflabel{\large$3$}{i3}
		\fmflabel{\large$4$}{i4}
		\fmflabel{\large$5$}{i5}
		\fmflabel{\large$6$}{i6}
		\fmflabel{\large$1'$}{o1}
		\fmflabel{\large$2'$}{o2}
		\fmflabel{\large$3'$}{o3}
		\fmflabel{\large$4'$}{o4}
		\fmflabel{\large$5'$}{o5}
		\fmflabel{\large$6'$}{o6}
		\end{fmfgraph*}
	\end{fmffile}
	} 
	\hspace*{\fill}
	\subfloat[Glued diagram resulting from the combination of Fig.~\ref{fig:disconnected-unperm} and Fig.~\ref{fig:disconnected-perm2}. \label{fig:glued-case2}]{
	\begin{fmffile}{fgraphs14}
		\begin{fmfgraph*}(50,50)
		\fmfset{thick}{1.25}
		\fmfpen{thick}
		\fmfforce{(.1w,.90h)}{i1} 
		\fmfforce{(.3w,.65h)}{i2} 
		\fmfforce{(.3w,.40h)}{o2} 
		\fmfforce{(.1w,.15h)}{o1}
		\fmfforce{(.9w,.90h)}{o5} 
		\fmfforce{(.7w,.65h)}{i3} 
		\fmfforce{(.7w,.40h)}{o4} 
		\fmfforce{(.9w,.15h)}{i6} 
		\fmf{plain, tension=1}{i1,v1}
		\fmf{plain, tension=1}{i2,v1}
		\fmf{plain, tension=1}{v1,o1}
		\fmf{plain, tension=1}{v1,o2}
		\fmf{plain, tension=1}{i3,v2}
		\fmf{plain, tension=1}{i6,v2}
		\fmf{plain, tension=1}{v2,o4}
		\fmf{plain, tension=1}{v2,o5}
		\fmf{dashes, tension=1}{i1,o1}
		\fmf{dashes, tension=1}{i3,o3}
		\fmf{dashes, tension=1}{o3,i2}
		\fmf{dashes, tension=1}{o2,i4}
		\fmf{dashes, tension=1}{i4,o4}
		\fmf{dashes, tension=1}{i6,o6}
		\fmf{dashes, tension=1}{o6,i5}
		\fmf{dashes, tension=1}{i5,o5}
		\fmfdot{v1}
		\fmfdot{v2}
		\fmfv{decor.shape=square,decor.filled=full,decor.size=2.5thick}{i1}
		\fmfv{decor.shape=square,decor.filled=full,decor.size=2.5thick}{i2}
		\fmfv{decor.shape=square,decor.filled=full,decor.size=2.5thick}{i3}
		\fmfv{decor.shape=square,decor.filled=full,decor.size=2.5thick}{i4}
		\fmfv{decor.shape=square,decor.filled=full,decor.size=2.5thick}{i5}
		\fmfv{decor.shape=square,decor.filled=full,decor.size=2.5thick}{i6}
		\fmfv{decor.shape=square,decor.filled=full,decor.size=2.5thick}{o1}
		\fmfv{decor.shape=square,decor.filled=full,decor.size=2.5thick}{o2}
		\fmfv{decor.shape=square,decor.filled=full,decor.size=2.5thick}{o3}
		\fmfv{decor.shape=square,decor.filled=full,decor.size=2.5thick}{o4}
		\fmfv{decor.shape=square,decor.filled=full,decor.size=2.5thick}{o5}
		\fmfv{decor.shape=square,decor.filled=full,decor.size=2.5thick}{o6}
		\fmflabel{\large$1$}{i1}
		\fmflabel{\large$2$}{i2}
		\fmflabel{\large$3$}{i3}
		\fmflabel{\large$4$}{i4}
		\fmflabel{\large$5$}{i5}
		\fmflabel{\large$6$}{i6}
		\fmflabel{\large$1'$}{o1}
		\fmflabel{\large$2'$}{o2}
		\fmflabel{\large$3'$}{o3}
		\fmflabel{\large$4'$}{o4}
		\fmflabel{\large$5'$}{o5}
		\fmflabel{\large$6'$}{o6}
		\end{fmfgraph*}
	\end{fmffile}
	} 
	\hspace*{\fill}
	\subfloat[Glued diagram resulting from the combination of Fig.~\ref{fig:disconnected-unperm} and Fig.~\ref{fig:disconnected-perm3}. \label{fig:glued-case3}]{
	\begin{fmffile}{fgraphs15}
		\begin{fmfgraph*}(50,50)
		\fmfset{thick}{1.25}
		\fmfpen{thick}
		\fmfforce{(.1w,.90h)}{i1} 
		\fmfforce{(.3w,.65h)}{i2} 
		\fmfforce{(.3w,.40h)}{o1} 
		\fmfforce{(.1w,.15h)}{o2}
		\fmfforce{(.9w,.90h)}{i3} 
		\fmfforce{(.7w,.65h)}{i6} 
		\fmfforce{(.7w,.40h)}{o4} 
		\fmfforce{(.9w,.15h)}{o5} 
		\fmf{plain, tension=1}{i1,v1}
		\fmf{plain, tension=1}{i2,v1}
		\fmf{plain, tension=1}{v1,o1}
		\fmf{plain, tension=1}{v1,o2}
		\fmf{plain, tension=1}{i3,v2}
		\fmf{plain, tension=1}{i6,v2}
		\fmf{plain, tension=1}{v2,o4}
		\fmf{plain, tension=1}{v2,o5}
		\fmf{dashes, tension=1}{i1,o3}
		\fmf{dashes, tension=1}{o3,i3}
		\fmf{dashes, tension=1}{o1,i4}
		\fmf{dashes, tension=1}{i4,o4}
		\fmf{dashes, tension=1}{i2,o6}
		\fmf{dashes, tension=1}{o6,i6}
		\fmf{dashes, tension=1}{o2,i5}
		\fmf{dashes, tension=1}{i5,o5}
		\fmfdot{v1}
		\fmfdot{v2}
		\fmfv{decor.shape=square,decor.filled=full,decor.size=2.5thick}{i1}
		\fmfv{decor.shape=square,decor.filled=full,decor.size=2.5thick}{i2}
		\fmfv{decor.shape=square,decor.filled=full,decor.size=2.5thick}{i3}
		\fmfv{decor.shape=square,decor.filled=full,decor.size=2.5thick}{i4}
		\fmfv{decor.shape=square,decor.filled=full,decor.size=2.5thick}{i5}
		\fmfv{decor.shape=square,decor.filled=full,decor.size=2.5thick}{i6}
		\fmfv{decor.shape=square,decor.filled=full,decor.size=2.5thick}{o1}
		\fmfv{decor.shape=square,decor.filled=full,decor.size=2.5thick}{o2}
		\fmfv{decor.shape=square,decor.filled=full,decor.size=2.5thick}{o3}
		\fmfv{decor.shape=square,decor.filled=full,decor.size=2.5thick}{o4}
		\fmfv{decor.shape=square,decor.filled=full,decor.size=2.5thick}{o5}
		\fmfv{decor.shape=square,decor.filled=full,decor.size=2.5thick}{o6}
		\fmflabel{\large$1$}{i1}
		\fmflabel{\large$2$}{i2}
		\fmflabel{\large$3$}{i3}
		\fmflabel{\large$4$}{i4}
		\fmflabel{\large$5$}{i5}
		\fmflabel{\large$6$}{i6}
		\fmflabel{\large$1'$}{o1}
		\fmflabel{\large$2'$}{o2}
		\fmflabel{\large$3'$}{o3}
		\fmflabel{\large$4'$}{o4}
		\fmflabel{\large$5'$}{o5}
		\fmflabel{\large$6'$}{o6}
		\end{fmfgraph*}
	\end{fmffile}
	}
	\newline
	\caption{Glued diagrams generated by identifying the field sources of one Feynman diagram in Fig.~\ref{fig:disconnected-perms} with the field sources of another and then connecting the diagrams. For visual clarity, propagators which are not part of the four-point function are represented with dashed lines and field sources are denoted by a small square. Insertions of the background field $\chi_{\mathrm{cl}}$ are omitted. \label{fig:glued-diagrams}}
\end{figure*}
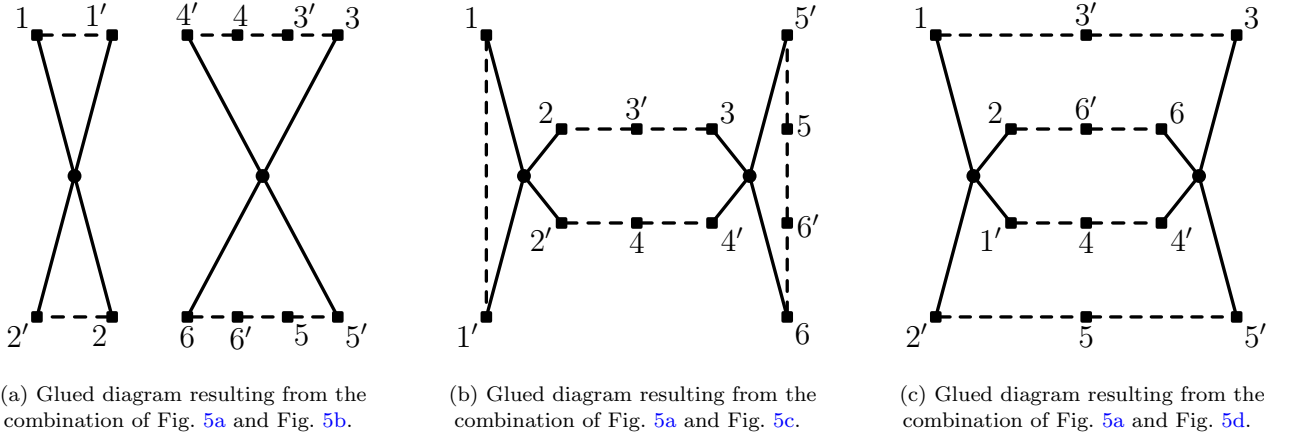

Deep properties of the permutations shown in Fig.~\ref{fig:disconnected-perms} which are not immediately apparent in the Feynman diagrams are made manifest in the ``glued diagrams'' of Fig.~\ref{fig:glued-diagrams}. For each type of permutation considered, the resulting graph topology in Fig.~\ref{fig:glued-diagrams} is different. The permutation of Fig.~\ref{fig:disconnected-perm1} results in the glued diagram of Fig.~\ref{fig:glued-case1} where the four-point interactions are in distinct subgraphs. This is in contrast to the permutations of Figs.~\ref{fig:disconnected-perm2} and~\ref{fig:disconnected-perm3} and their glued diagrams, Figs.~\ref{fig:glued-case2} and~\ref{fig:glued-case3}, which feature both four-point interactions in the same subgraph with a particular topology. In Fig. ~\ref{fig:glued-case2}, colloquially, two of the legs of each four-point interaction are connected to themselves resulting in ``capped ends''. Topologically, there exist closed cycles one can draw on the graph which pass through only one four-point function. In the case of Fig.~\ref{fig:glued-case3}, each leg of one four-point interaction is connected to a leg of the other four-point interaction. Topologically, this requires any closed cycle to pass through both four-point functions. The three topologies presented in Figs.~\ref{fig:glued-case1},~\ref{fig:glued-case2}, and~\ref{fig:glued-case3} are the only types possible for subgraphs constructed with exactly two four-point functions. As such, we can use these three possible topologies to categorize all possible permutations of Fig.~\ref{fig:disconnected-unperm}.

While the topological approach of the ``glued'' diagrams has already proven useful in catagorizing permutations, its greatest value is in making the sum over particle number in Eq.~\eqref{eq:Work-Dist-Scattering} tractable. While left implicit in Eq.~\eqref{eq:perms-final}, the sum over particle number is troublesome because it requires one to first find a closed form expression for the $n \rightarrow n$ work distribution function in terms of the particle number $n$ and then find a closed form expression for the infinite sum. This must be done in such a manner that the cancellation of the potentially divergent normalization factor $\mrm{tr}\{ \exp{(-\beta \hat{H}_{0})}\}= \exp{(-\beta F_{0})}$ is manifest. The glued diagram approach has the advantage of rephrasing the sum over particle number in terms of a sum over certain simple properties of the glued diagram.

The actual mathematical manipulations that go into the procedure are tedious and uninformative but a high level description is provided here instead. Consider the Feynman diagram in Fig.~\ref{fig:disconnected-perm3} and the glued diagram Fig.~\ref{fig:glued-case3}. For $n > 6$, Fig.~\ref{fig:disconnected-perm3} will include additional field sources and propagators. In the glued diagram, these propagators will either enter the subgraph containing the four-point functions, lengthening the paths in the graph but not changing the topology, or will create cycle graphs made entirely of propagators. With appropriate combinatorial factors, the sum over particle number can then be rephrased in terms of a sum over the length of paths in the subdiagram of the four-point functions, a sum over the number of disconnected cycles of propagators, and a sum over the length of each of these cycles. Importantly, the sum over the length of paths in the subgraph Fig.~\ref{fig:glued-case3} is independent of the sums over the number and length of cycles of disconnected propagators. This sum over disconnected cycles of propagators is just the sum over all possible ``trivial'' scatterings where the four-point function never appears. Carrying out this sum ultimately yields the normalization factor $\mrm{tr}\{ \exp{(-\beta \hat{H}_{0})}\}$. One may then evaluate the sum over path lengths in Fig.~\ref{fig:glued-case3} by noting that each propagator is a delta-function and each incoming field source is associated with a Boltzmann weight $\exp{(-\beta \omega)}$. This gives a set of geometric series which can be summed into Bose-Einstein statistics factors. While not shown here, it can be demonstrated that, due to their subgraph topology, Figs.~\ref{fig:glued-case1} and~\ref{fig:glued-case2} are proportional to $\delta\left(W\right)$ and thus make trivial contributions to the work distribution function. While the exact diagrammatics differ, this scheme applies equally well to $n \rightarrow n \pm 2$ and $n \rightarrow n \pm 4$ processes.

\section{Work Distribution Functions}
\label{app:Work-Distributions}
In the calculation of the work distribution function Eq.~\eqref{eq:Work-Dist-Scattering} for the time-dependent field theory \eqref{eq:Time-dependent-EFT}, it was found that the distribution factored into five distinct parts. These correspond to realizations of the experiment where the particle number is unchanged, the particle number increases or decreases by two, or the particle number increases or decreases by four. These are denoted by the distributions $\rho_{n \rightarrow n} \left( W \right)$, $\rho_{n \rightarrow n \pm 2} \left( W \right)$, and $\rho_{n \rightarrow n \pm 4} \left( W \right)$ respectively. As explained in Sec.~\ref{sec:Generic-Protocol}, these distributions are not normalized and this must be done by hand.

The work distribution functions for when the particle number is constant or changes by four are given by
\begin{align}
\rho_{n \rightarrow n - 4} \left( W \right) = & \frac{1}{4!} V \left| \int dt \, \lambda(t) \e{i W t} \right|^2 \int \widetilde{d^{3}k_{1}} \, \widetilde{d^{3}k_{2}} \, \widetilde{d^{3}k_{3}} \, \widetilde{d^{3}k_{4}} \; \delta \left( W + \omega_{1} + \omega_{2} + \omega_{3} + \omega_{4} \right) \, \left( 2 \pi \right)^3 \delta^3 \left( k_1 + k_2 + k_3 + k_4 \right) \nonumber \\
& \times \left( \frac{1}{\e{\beta \omega_1} - 1} \right) \left( \frac{1}{\e{\beta \omega_2} - 1} \right) \left( \frac{1}{\e{\beta \omega_3} - 1} \right) \left( \frac{1}{\e{\beta \omega_4} - 1} \right) , \label{eq:Work-Dist-Nm4-app}\\
\rho_{n \rightarrow n} \left( W \right) =& \frac{1}{4} V \left| \int dt \, \lambda(t) \e{i W t} \right|^2 \int \widetilde{d^{3}k_{1}} \, \widetilde{d^{3}k_{2}} \, \widetilde{d^{3}k_{1}'} \, \widetilde{d^{3}k_{2}'} \; \delta \left( W + \omega_{1} + \omega_{2} - \omega_{1}' -\omega_{2}' \right) \, \left( 2 \pi \right)^3 \delta^3 \left( k_1 + k_2 - k_1' -k_2' \right) \nonumber \\
& \times \left( \frac{1}{\e{\beta \omega_1} - 1} \right) \left( \frac{1}{\e{\beta \omega_2} - 1} \right) \left( 1 + \frac{1}{\e{\beta \omega_1'} - 1} \right) \left( 1 + \frac{1}{\e{\beta \omega_2'} - 1} \right) , \label{eq:Work-Dist-N-app} \\
\rho_{n \rightarrow n + 4} \left( W \right) = & \frac{1}{4!} V \left| \int dt \, \lambda(t) \e{i W t} \right|^2 \int \widetilde{d^{3}k_{1}} \, \widetilde{d^{3}k_{2}} \, \widetilde{d^{3}k_{3}} \, \widetilde{d^{3}k_{4}} \; \delta \left( W - \omega_{1} - \omega_{2} - \omega_{3} - \omega_{4} \right) \, \left( 2 \pi \right)^3 \delta^3 \left( k_1 + k_2 + k_3 + k_4 \right) \nonumber \\
& \times \left( 1 + \frac{1}{\e{\beta \omega_1} - 1} \right) \left( 1 + \frac{1}{\e{\beta \omega_2} - 1} \right) \left( 1 + \frac{1}{\e{\beta \omega_3} - 1} \right) \left( 1 + \frac{1}{\e{\beta \omega_4} - 1} \right) . \label{eq:Work-Dist-Np4-app}
\end{align}
In the calculation of these work distributions only tree-level diagrams enter. Thus, even though particle number is not conserved, these work distributions can be calculated from the classical equations of motion. This should be contrasted with the work distributions for when the particle number changes by two. Loop diagrams contribute to these work distributions and their contributions can be separated out,
\be
\rho_{n \rightarrow n\pm 2} \left( W \right) = \rho^{\textrm{tree}}_{n \rightarrow n \pm 2} \left( W \right) + \rho^{\textrm{loop}}_{n \rightarrow n \pm 2} \left( W \right).
\ee
The tree-level contributions to the work distributions are given by
\begin{align}
\rho^{\textrm{tree}}_{n \rightarrow n + 2} \left( W \right) = & \frac{1}{3!} V \left| \int dt \, \lambda(t) \e{i W t} \right|^2 \int \widetilde{d^{3}k_{1}} \, \widetilde{d^{3}k_{1}'} \, \widetilde{d^{3}k_{2}'} \, \widetilde{d^{3}k_{3}'} \; \delta \left( W + \omega_{1} - \omega_{1}' - \omega_{2}' -\omega_{3}' \right) \, \left( 2 \pi \right)^3 \delta^3 \left( k_1 - k_1' - k_2' - k_3' \right) \nonumber \\
& \times \left( \frac{1}{\e{\beta \omega_1} - 1} \right) \left( 1 + \frac{1}{\e{\beta \omega_1'} - 1} \right) \left( 1 + \frac{1}{\e{\beta \omega_2'} - 1} \right) \left( 1 + \frac{1}{\e{\beta \omega_3'} - 1} \right) , \\
\rho^{\textrm{tree}}_{n \rightarrow n - 2} \left( W \right) = & \frac{1}{3!} V \left| \int dt \, \lambda(t) \e{i W t} \right|^2 \int \widetilde{d^{3}k_{1}} \, \widetilde{d^{3}k_{2}} \, \widetilde{d^{3}k_{3}} \, \widetilde{d^{3}k_{1}'} \; \delta \left( W + \omega_{1} + \omega_{2} + \omega_{3} - \omega_{1}' \right) \, \left( 2 \pi \right)^3 \delta^3 \left( k_1 + k_2 + k_3 - k_1' \right) \nonumber \\
& \times \left( \frac{1}{\e{\beta \omega_1} - 1} \right) \left( \frac{1}{\e{\beta \omega_2} - 1} \right) \left( \frac{1}{\e{\beta \omega_3} - 1} \right) \left( 1 + \frac{1}{\e{\beta \omega_1'} - 1} \right) .
\end{align}
These tree-level contributions follow the same pattern as the work distributions \eqref{eq:Work-Dist-Nm4-app}-\eqref{eq:Work-Dist-Np4-app}. The contributions which arise from the loop diagrams are
\begin{align}
\rho^{\textrm{loop}}_{n \rightarrow n + 2} \left( W \right) = & \frac{1}{2} V \left| \int dt \, \lambda(t) \e{i W t} \right|^2 \left( 1 + \frac{1}{\e{\beta W / 2} - 1} \right)^2 \frac{1}{W} \left( \int \widetilde{d^{3}k} \, \delta \left( W - 2 \omega \right) \right) \nonumber \\
& \times \left( \int \widetilde{d^{3}k} \, \frac{1}{\e{\beta \omega} - 1} + \frac{1}{2} \left( \frac{m}{4 \pi} \right)^2 \left[ 1 + \log \left( \frac{\mu^2}{m^2} \right) \right] \right)^2 , \\
\rho^{\textrm{loop}}_{n \rightarrow n - 2} \left( W \right) = & \frac{1}{2} V \left| \int dt \, \lambda(t) \e{i W t} \right|^2 \left( \frac{1}{\e{- \beta W / 2} - 1} \right)^2 \frac{1}{-W} \left( \int \widetilde{d^{3}k} \, \delta \left( W + 2 \omega \right) \right) \nonumber \\
& \times \left( \int \widetilde{d^{3}k} \, \frac{1}{\e{\beta \omega} - 1} + \frac{1}{2} \left( \frac{m}{4 \pi} \right)^2 \left[ 1 + \log \left( \frac{\mu^2}{m^2} \right) \right] \right)^2 .
\end{align}
In these expressions, $\mu$ is the $\overline{MS}$ renormalization scale. These expressions do partially include contributions from tree-level diagrams because the loop and tree diagrams interfere.
\end{widetext}
\end{appendix}


\bibliography{QFT-J}


\end{document}